\documentclass[aps,prd]{revtex4-2}
\usepackage[pdftex]{graphicx}
\usepackage{amssymb}
\usepackage{amsmath}
\usepackage{orcidlink}
\usepackage{bm}
\usepackage{hyperref}

\begin{document}
\setcounter{page}{1}
\title[]{Determination of the Angular Momentum  of Radiated Gravitons, Scalars, and Dark Photons from Binary Orbits}
\author{Arpan Hait\orcidlink{0000-0001-7540-0111}}
\email{arpanhait867@gmail.com}
\thanks{(Corresponding author)}
\affiliation{Department of Physics, Indian Institute of Technology Kanpur, Kanpur, 208016, India}
\author{Subhendra Mohanty\orcidlink{0000-0003-0070-6647}}
\email{subhendram@iiserb.ac.in}
\affiliation{Indian Institute of Science Education and Research Bhopal, Bhopal, 462066, India}
\date[]{Dated: \today}

\begin{abstract}
 We compute the energy and angular momentum radiated by compact binaries through gravitational waves, scalar and vector fields, using the field-theoretic approach. We observe that each emitted graviton carries away an angular momentum of $\dot{J}/\dot {N}\simeq 2\hbar$, not only for elliptical orbits (as has already been pointed out by Page~\cite{Page:2025ncu}) but also for hyperbolic orbits.  We extend the analysis to scalar and vector radiation and find that the angular momentum carried by each emitted quantum of radiation is approximately $\hbar$, i.e., $\dot{J}/\dot{N}\simeq \hbar$ for both types of radiation. We demonstrate that the radiated energy and angular momentum of binary systems can be determined from the evolution of the orbital angular frequency and the eccentricity of the quasi-Kepler orbit. However, a gauge-invariant decomposition of the angular momentum carried by the emitted particles into the spin and orbital components is not possible. 
\end{abstract}


\maketitle

\section{INTRODUCTION}


Ultralight fields with masses in the range of $m\sim 10^{-19} {\rm eV}$ and a Compton-wavelength of a few ${\rm kpc}$ can be viable candidates for the ``fuzzy dark matter"~\cite{Hu:2000ke, Hui:2016ltb} which can solve some of the problems of the standard cold-dark matter, like the core-cusp problem and missing satellite problem~\cite{Schive:2025bcm}. It is an interesting coincidence that the time-period of compact-binary pulsars like the Hulse-Taylor binary~\cite{Hulse:1974eb}  $T\sim 8\, {\rm hrs}$ corresponds to the angular frequency $\omega\sim 2\pi/P\sim 10^{-19} \,{\rm }eV$ which is in the range of fuzzy-dark matter mass. Monitoring the dynamics of the Hulse-Taylor binary pulsar and other compact binary systems was the first indirect test of the existence of gravitational waves~\cite{Weisberg:2010zz}. The time-period of the binary orbits decreases at the rate $\dot{P} \sim 10^{-12} ~s/s$, which agrees with the prediction from the calculation of period loss expected from binaries in elliptical orbits due to gravitational waves by Peters and Mathews~\cite{Peters:1963ux}. Hulse-Taylor type binaries can also radiate scalar and vector particles as long as their masses are smaller than the orbital frequency. Therefore, compact binary dynamics is a good probe of ultralight scalar~\cite{Mohanty:1994yi, KumarPoddar:2019jxe, Poddar:2023pfj}  or vector~\cite{KumarPoddar:2019ceq} radiation, as there will be additional channels of energy loss of binary orbits due to radiation of scalar and vector particles in addition to gravitational waves. The decrease in the total energy of the binary due to the radiation of gravitational waves and other ultralight particles reduces the relative distance and the orbital period, as observed experimentally. This observation is then used to constrain the couplings and masses of the possible extra long-range fields associated with the neutron star, white dwarf, or black holes in the binary. For binaries in elliptical orbits, there is a loss of the binary's total angular momentum, resulting in a decrease in orbital eccentricity, which can be observed. The total angular momentum carried away by gravitational waves from binaries was calculated by Peters~\cite{Peters:1964zz}. 

Recently, Page~\cite{Page:2025ncu} used the Peters-Mathews formula for energy radiated~\cite{Peters:1963ux} and the Peters formula~\cite{Peters:1964zz} for total angular momentum radiated due to gravitational waves, and pointed out a curious fact that the ratio of the angular momentum radiated in gravitational radiation with the number of gravitons, i.e., $\dot{J}/\dot{N}$ is approximately equal to $2\hbar$. Here, the graviton number is a secondary quantity derived from the primary observable, the rate of energy radiated as $\dot{N} \equiv \dot{E}/ (\hbar \omega_0)$, where $\omega_0= \sqrt{GM/a^3}$ is the fundamental angular frequency of the elliptical orbit. 
 
 In this paper, we examine the angular momentum carried away by scalar, vector, and graviton fields. We compute the total angular momentum and energy carried away by fields of different spins from binary stars in elliptical and hyperbolic orbits. We perform the calculation using the field-theoretic method, in which the spectra of the angular momentum and radiated energy are derived in the frequency domain.   This method enables us to compute the ratio $\dot{J}(\omega)/\dot{N}(\omega)$ for a given frequency. For graviton, we find that this ratio deviates significantly from $2\hbar$ for the lower multipoles $n \omega_{0}$ of the fundamental frequency and saturates to $2\hbar $ at large $n$. 

The loss of total angular momentum can be observed by measuring the change in eccentricity of the elliptical binary orbit with time~\cite{Peters:1964zz}. The total angular momentum can be defined in terms of momentum and position operators for spin-2~\cite{Maggiore:2007ulw} and spin-1~\cite{Manohar:2022dea} fields.  Using these definitions, we compute the total angular-momentum radiation formulas in the field-theoretic method to compare with observations. The total angular momentum $\vec{J}= \vec{L} + \vec {S}$, which is the sum of the orbital angular momentum $\vec{L}$ and the spin $\vec{S}$, is a gauge invariant quantity. However neither $\vec L$ nor $\vec S$ are individually gauge invariant either for spin-2~\cite{Maggiore:2007ulw} or spin-1~\cite{Manohar:2022dea} fields. Hence, from the observations of binary orbits, we can determine the total angular momentum loss. Still, we cannot uniquely split this in a gauge-invariant way into spin and orbital parts. The same calculation is carried out for ultralight scalars radiated from compact binaries. For the case of scalar radiation, it is found that the ratio of the total angular momentum radiated to particle number $\dot{J}/\dot{N} \simeq \hbar$, which is the same as that for vector radiation. This reflects the dipolar nature of the source for both cases. We conclude from these studies that studying the angular momentum radiated from elliptic and hyperbolic binaries can be a useful way to determine whether the energy loss is also due to ultralight scalars or vector particles.  The ratio $\dot{J}/\dot {N} \simeq 2\hbar $ for gravitons can be attributed to the quadrupole distribution of the source. From the orbital dynamics of binaries, one cannot prove that the spin of the gravitons is quantized, as only the total spin plus orbital angular momentum is observed from orbital dynamics.

The paper is organized as follows: In Sec.~\ref{Sec:classical_background}, we outline the framework for computing the energy and angular momentum losses due to gravitational radiation in the field-theoretic approach. We evaluate the angular momentum radiated per quantum for graviton emission from elliptic orbits and hyperbolic encounters in Sec.~\ref{Sec:Elliptic_Gravity} and Sec.~\ref{Sec:Hyperbolic_Gravity}, respectively. The angular momentum loss per radiated quantum for non-gravitational interaction is obtained in Sec.~\ref{Sec:non_gravity_rad}. In Sec.~\ref{Sec:spin_determination}, we discuss how the spin of the emitted quanta can be determined from the binary observations. Finally, we summarize our conclusion in Sec.~\ref{Sec:conclusions}.

\section{Graviton radiation from classical sources}\label{Sec:classical_background}

\noindent We define the mass dimension one graviton field $h_{\mu \nu}$ as the deviation of the metric with respect to the Minkowski space, 
\begin{eqnarray}
g_{\mu \nu} = \eta_{\mu \nu} + \kappa h_{\mu \nu},
\end{eqnarray}
 where $\kappa = \sqrt{32\pi G}$, which makes the graviton kinetic term be in the canonical form. The dimensionless strain measured by gravitational wave detectors is $\kappa h_{\mu\nu}$. Expanding the matter Lagrangian in curved space to the linear order in $\kappa h_{\mu \nu}$ then gives us the interaction vertex between matter and gravitons as
\begin{eqnarray}
    {\mathcal L}_{\rm int} = \frac{\partial L_{m}}{\partial  g_{\mu \nu}}\Bigg\vert_{g_{\mu\nu} =\eta_{\mu\nu}}
\,\kappa h_{\mu \nu}= \frac{\kappa}{2}h_{\mu \nu}T^{\mu \nu},
\end{eqnarray}
which implies that gravitons couple to matter (and gravitons) through the stress tensor with a coupling of $\kappa/2$. We will treat the source current $T^{\mu\nu}$ as classical, and gravitons as quantum fields,
\begin{eqnarray}
\hat {h}_{\mu \nu}(x) = \sum_\lambda \int \frac{d^3 k}{(2 \pi)^3}\,\sqrt{\frac{\hbar}{2 \omega_k}}\, \left[\epsilon^\lambda_{\mu\nu}(k)\,a_\lambda(k)\,e^{-\iota k \cdot x} + \epsilon^{*\lambda}_{\mu \nu}(k)\, a^\dagger_\lambda(k)\,e^{\iota k \cdot x}  \right].
\label{graviton}
\end{eqnarray}
The graviton creation and annihilation operators obey the usual bosonic commutation relations $[a_\lambda(k),a^\dagger_{\lambda^\prime}(k^\prime)]=(2\pi)^3\,\delta_{\lambda \lambda^\prime}\,\delta^3(\vec k -\vec k^\prime)$. The $\hbar$ dependence in Eq.~\eqref{graviton} ensures that the commutation relation between the graviton field $\hat{h}_{ij}(\vec x,t)$ and its conjugate momenta $\hat{\pi}^{lm}=(\dot{h}^{lm}-\dot{h} \delta^{lm})$ has the canonical form, $[\hat h_{ij}(\vec x,t)
, \hat \pi^{lm}(\vec y,t)] = \iota\hbar\, \delta^l_{[i}\delta^m_{j]}\, \delta^3(\vec x-\vec y)$. We choose units where $c=1$ but retain $\hbar$ to keep track of the quantum nature of the processes.     
Here $\epsilon^\lambda_{\mu\nu}(k)$ are the polarization tensors which obey the orthogonality 
 $\epsilon^\lambda_{\mu  nu}(k) {\epsilon^{*\lambda^\prime}}^{\mu \nu}(k)= \delta_{\lambda \lambda^\prime}$
and completeness relations,
\begin{eqnarray}\label{comp}
    \sum^{2}_{\lambda\,=\,1} \epsilon^{\lambda}_{\mu\nu} (k) \epsilon^{\lambda}_{ \alpha\beta} (k) \equiv P_{\mu\nu, \alpha\beta} = \cfrac{1}{2} \left(\eta_{\mu \alpha} \eta_{\nu \beta} + \eta_{\mu \beta} \eta_{\nu \alpha} - \eta_{\mu \nu} \eta_{\alpha \beta}\right).
\end{eqnarray}
The `in' state is the vacuum state $|0\rangle$ with zero gravitons, and gives the `out' state
\begin{equation}
|{\rm{out}} \rangle = e^{(\iota/\hbar) \int \widetilde{dk}\, \frac{\kappa}{2}h_{\mu \nu}T^{\mu \nu}(k)} |0\rangle\,\equiv \left(1 + \iota \mathcal{T}\right)|0\rangle,
\end{equation}
where $
\widetilde{dk} = \cfrac{d^{3}k}{\left(2\pi\right)^3 2\omega_{k}}$ is the phase space for each outgoing graviton. When we have a classical source $T_{\mu \nu}$ and gravitons are quantum fields as in Eq.~\eqref{graviton}, the `out' state of gravitons will be a coherent state~\cite{Britto:2021pud, Kanno:2025how} which is a sum of different occupation number states. In this work, we consider periodic sources and compute the average values of the graviton, energy, and angular momentum fluxes, all averaged over one cycle of the source's orbit. The average value of the operator $\mathcal{O}$ in the `out' state can be expressed as
\begin{eqnarray}
    \langle  \rm{out}\, | {\mathcal{O}}| \rm{out} \rangle= \langle 0 | \mathcal{T}^{\dagger} {\mathcal{O} } \mathcal{T} |0\rangle.
\end{eqnarray}
To the leading order in $\kappa$, the probability per unit time of radiation of gravitons with energy $\hbar\omega_k$ from classical sources is then, analogous to the formula for radiation of photons from classical electromagnetic currents, and is given by~\cite{Heitler:1936jqw, harris2014pedestrian},
\begin{eqnarray}
    d\Gamma = \cfrac{2\pi}{\hbar} \,\delta( E^\prime- \hbar\omega_k)\, \cfrac{\kappa^2}{4}\,\sum_\lambda \hbar\,|\epsilon^{ij}_{\lambda \rm TT}(k)\,T_{i j}(k^\prime)|^2\,\cfrac {d^3k}{(2\pi)^3 2 \omega_k }\,,
    \label{dgamma}
\end{eqnarray}
where $E^\prime$ is the energy associated with the angular frequency modes of the classical current $T_{ij}(k^\prime)$ and $\hbar\omega_k=\hbar k$ is the energy of gravitons with wavenumber of magnitude $k=|\vec k|$. Here we have considered the emission of the transverse-traceless component of the graviton, which is obtained by taking the TT projection of $\epsilon^{\mu \nu}_{\lambda}$ with respect to the direction of propagation $\hat{k}$ of the graviton,
\begin{eqnarray}\label{TTP} 
   \epsilon^{\rm TT}_{\lambda ij} (k)= {\Lambda_{ij, lm}}(\hat k)\, \epsilon_{\lambda lm }(k),
\end{eqnarray}
such that $\hat k_j {\epsilon^{\rm TT}_{\lambda}}_{ij}(k)=\hat k_i {\epsilon^{\rm TT}_{\lambda}}_{ij}(k)=0$. The TT projection operator depends on $\hat k$ and is given by~\cite{Weinberg:1972kfs, Maggiore:2007ulw},
\begin{eqnarray}\label{TTPk}
    \Lambda_{ij, lm}(\hat k) &=& \left(\delta_{il} - \hat{k}_{i}\hat{k}_{l}\right)\left(\delta_{jm} -\hat{k}_{j}\hat{k}_{m}\right)-\cfrac{1}{2}\left(\delta_{ij}-\hat{k}_{i}\hat{k}_{j}\right)\left(\delta_{lm}-\hat{k}_{l}\hat{k}_{m}\right)\nonumber \\
    &=&\left[\delta_{il}\,\delta_{jm} -\delta_{il}\,\hat{k}_{j}\hat{k}_{m}-\delta_{jm}\,\hat{k}_{i}\hat{k}_{l} -\cfrac{1}{2}\delta_{ij}\,\delta_{lm} +\cfrac{1}{2}\delta_{ij}\,\hat{k}_{l}\hat{k}_{m} + \cfrac{1}{2}\delta_{lm}\,\hat{k}_{i}\hat{k}_{j} + \cfrac{1}{2}\hat{k}_{i}\hat{k}_{j}\hat{k}_{l}\hat{k}_{m}\right].
\end{eqnarray}
The amplitude squared of the graviton emission process from a classical source $T_{ij}(k^\prime)$ is
\begin{eqnarray}
    |{\mathcal{M}}|^2= \frac{\kappa^2}{4}\,\hbar\sum_\lambda \epsilon_{\lambda \rm TT}^{*i j}(k) T_{i j}^*(k^\prime)\epsilon_{\lambda \rm TT}^{lm}(k) T_{lm}(k^\prime),
\end{eqnarray}
which using the completeness relation in Eq.~\eqref{comp}, the TT projection relation in Eq.~\eqref{TTP} and the property of the projection operator, $\Lambda_{ij, lm}(\hat k)\,\Lambda_{lm, pq}(\hat k)={\Lambda_{ij, pq}} (\hat k)$ reduces to
\begin{eqnarray}
    |{\cal M}|^2= \frac{\kappa^2}{4}\,\hbar {\Lambda_{ij, lm}(\hat k)} T^{*ij}(k^\prime) T^{lm}(k^\prime).
\end{eqnarray}
The stress tensor in momentum space $T^{lm}(k^\prime)$ is taken to be the Fourier transform of the position space stress tensor, 
\begin{eqnarray}
T_{ij}(\textbf{k}^\prime,\omega^\prime)&=&\frac{1}{T}\int_0^T dt^\prime \int d^3x^\prime\,\tilde{T}_{ij}(\textbf{x}^\prime,t^\prime)\,e^{-\iota(\textbf{k}^\prime\cdot \textbf{x}^\prime-\omega^\prime t^\prime)}\nonumber\\
&=& \int \tilde{T}_{ij}(\textbf{x}^\prime,\omega^{\prime})\, e^{-\iota\textbf{k}^\prime\cdot \textbf{x}^\prime}\,d^3x^\prime\simeq \int \tilde{T}_{ij}(\textbf{x}^\prime,\omega^{\prime})\,d^3x^\prime.
\label{Tij} 
\end{eqnarray}
Here we have made the quadrupole approximation $e^{-\iota\textbf{k}^\prime\cdot \textbf{x}^\prime}\sim 1$, which is a good assumption when the size of the source is much smaller than the wavelength of the gravitational wave, $|\textbf{x}^\prime|\ll 1/|\textbf{k}^\prime|= 1/\omega_k =\lambda_{gw}/(2\pi)$. The frequency of the gravitational waves emitted by a time-varying source will be related to the harmonics of the source frequency, so the quadrupole approximation is valid for binaries in compact orbits moving with non-relativistic speeds. In the quadrupole approximation the stress tensor $T_{\mu \nu}(\omega^\prime, \mathbf{k}^\prime)$ is independent of $\mathbf{k}^\prime$. In general (for higher multipoles), the relation of the radiation field to the source is via the Green's function relation, which in momentum space is
\begin{eqnarray}
    h_{\mu \nu}(k) = \int d^4 k^\prime\, \delta^{4} (k-k^\prime) G_{\mu\nu, \alpha \beta}(k) T_{\alpha \beta}(k^\prime).
\end{eqnarray}
Hence, they have in general for the source current momenta $\mathbf{k}^\prime=\omega_{k} \hat{\,\mathbf k}$ and the source frequency $\omega^\prime=\omega_k$. From Eq.~\eqref{dgamma}, we can write the rate of energy $\hbar\omega_k$ radiated as,
\begin{eqnarray}\label{dotE1}
    \cfrac{dE}{dt} = \cfrac{\kappa^2}{4} \int \cfrac{d^3 k}{\left(2\pi\right)^3 2\omega_{k}}\,{\Lambda}_{ij, lm} (\hat{k})T^{*ij} ({\omega}^\prime) (\hbar \omega_{k}) T^{lm}({\omega}^\prime) \left(2\pi\right) \delta(E^{\prime} - \hbar \omega_{k}).
\end{eqnarray}
For a classical source, we express the stress tensor in a Fourier series with frequency $\omega^{\prime} = n\omega_{0}$ with $\omega_{0}$ being the fundamental frequency. For example, for a elliptical binary $\omega_{0} = \sqrt{G M / a^3}$. We express the delta function in Eq.~\eqref{dotE1} in terms of the source frequency and graviton frequency $\delta(E^\prime-\hbar\omega_{k})=\hbar^{-1} \delta(\omega^\prime- \omega_{k}) $. The expression in Eq.~\eqref{dotE1} for the rate of energy radiated can be written as
\begin{eqnarray}\label{dotE2}
\cfrac{dE}{dt} 
    = \cfrac{\kappa^2}{32 \pi^2}\int d\omega_{k}d\Omega_{k} \delta( \omega^\prime- \omega_k) \omega^{2}_{k} {\Lambda_{ij,lm}(\hat k)}T^{*ij}(\omega^{\prime})T^{lm}(\omega^{\prime}).  
\end{eqnarray}
We see that the rate of energy radiated in forms of gravitons from a classical source $T_{ij}(\omega^\prime)$ is independent of $\hbar$. In the quadrupole approximation, the source terms $T_{ij}(\omega^\prime)$ have no $\hat{k}$ dependence, therefore the angular integral in Eq.~\eqref{dotE2} only involves ${\Lambda_{ij,\,lm}(\hat k)}$ as given in Eq.~\eqref{TTPk}. Using the relations~\cite{Weinberg:1972kfs,Maggiore:2007ulw}, 
\begin{equation}
\int d\Omega\, \hat{k}_{i}\hat{k}_{j}=\frac{4\pi}{3}\,\delta_{ij}, \qquad \int d\Omega \, \hat{k}_{i}\hat{k}_{j}\hat{k}_{l}\hat{k}_{m}=\frac{4\pi}{15}(\delta_{ij} \delta_{lm}+\delta_{il} \delta_{jm}+\delta_{im} \delta_{jl}),
\label{dOmegak1}
\end{equation}
we have 
\begin{eqnarray}
    \int d\Omega_k  {\Lambda_{ij, lm}(\hat k)}= \frac{2\pi}{15} \left( 11\, \delta_{il} \delta_{jm}- 4\, \delta_{ij} \delta_{lm} + \delta_{im} \delta_{jl} \right).
\end{eqnarray}
Which gives us
\begin{eqnarray}
\int d\Omega_k \Lambda_{ij, lm}(\hat k) T^{*ij}(\omega^\prime) T^{lm}({\omega^\prime}) = \frac{8\pi}{5}\left(T_{ij}(\omega^\prime) T^*_{ji}(\omega^\prime)-\frac{1}{3}\Big|T^{i}{}_{i}(\omega^\prime)\Big|^2\right).
\label{domegaTij}
\end{eqnarray}
Using the expression of Eq.~\eqref{domegaTij} in Eq.~\eqref{dgamma}, we obtain the rate of radiated graviton number 
\begin{eqnarray}\label{dNdt}
\cfrac{dN}{dt}=\cfrac{8G}{5\hbar}\int \left[T_{ij}(\omega^{\prime})T^*_{ji}(\omega^{\prime})-\frac{1}{3}|T^{i}{}_{i}(\omega^{\prime})|^2\right]
\delta(\omega^{\prime}-\omega_{k})\omega_{k} d\omega_{k},
\end{eqnarray}
and from Eq.~\eqref{dotE2}, we have the rate of energy radiated as gravitons,
\begin{eqnarray}\label{dEdt}
\cfrac{dE}{dt}=\cfrac{8G}{5}\int \left[T_{ij}(\omega^{\prime})T^*_{ji}(\omega^{\prime})-\frac{1}{3}|T^{i}{}_{i}(\omega^{\prime})|^2\right]
\delta(\omega^{\prime}-\omega_{k})\omega_{k}^{2} d\omega_{k}.
\end{eqnarray}
Next, we turn to the computation of the total angular momentum radiated through gravitons. The spin and the orbital angular momentum of gravitons can be defined as follows. Under spatial rotation $x^i\rightarrow x^i +R^{i}_{{}\,j}x^j$, the graviton field $h_{ij}(\textbf{x},t)$ transforms as
\begin{eqnarray}\label{hijrot}
    h_{ij}(\textbf{x},t) \rightarrow h_{ij}(\textbf{x},t) + \left( \omega^{lm}  (x_{l} \partial_{m} -x_{m} \partial_{l}) h_{ij}  + \omega_{i}^{{}\, l}  h_{lj} + \omega_{{}\,j}^{ l}  h_{il}  \right),
\end{eqnarray}
where the components of the antisymmetric tensor $\omega^{lm}$ are the three parameters of rotation. The generators of the two types of transformations in Eq.~\eqref{hijrot} are the orbital angular momentum operator $L_{lm}$ and spin angular momentum operators $S_{ij}$, which represent the angular momentum and spin three vectors of gravitons as $L^{a}= \epsilon^{alm} L_{lm}$ and $S^{a}= \epsilon^{alm} S_{lm}$, which can be written in terms of the graviton fields and their conjugate momenta as,
\begin{eqnarray}
    L^{a}= \epsilon^{alm} \int d^3 x\,  \pi_{ij}(\textbf{x},t) \left(  x_{l} \partial_{m} -x_{m} \partial_{l}\right)h_{ij}(\textbf{x},t),
\end{eqnarray}
and
\begin{eqnarray}
    S^a= \epsilon^{alm} \int d^3 x \, \pi_{lj}(\textbf{x},t)
h_{jm}(\textbf{x},t).
\end{eqnarray}
Treating $h_{ij}$ and $\pi_{ij}$ as quantum fields in Eq.~\eqref{graviton}, we can express $L^a$ and $S^a$ in terms of the momentum space creation and annihilation operators as,
\begin{eqnarray}\label{Lgr}
  \hat  L^a=-\iota \hbar \epsilon^{alm} \int \widetilde {dk}\, \sum_{\lambda}   \epsilon^{\lambda *}_{ij}(\hat k) a_\lambda ^\dagger(k) P^{ij,rs}(\hat k) \left( k_{l} \frac{\partial}{\partial k_{m}}\right)a_\lambda (k)  \epsilon^{\lambda }_{rs}(\hat k),
\end{eqnarray}
and 
\begin{eqnarray}\label{Sgr}
 \hat   S^a=-2\iota \hbar \epsilon^{alm} \int \widetilde{dk}\, \sum_{\lambda} a_{\lambda}^\dagger(k)  a_\lambda (k) \epsilon^{\lambda*}_{lj}(\hat k)\epsilon_{jm}^\lambda(\hat k),
\end{eqnarray}
respectively~\cite{Maggiore:2007ulw, Manohar:2022dea, DiVecchia:2022owy}. Under the general coordinate transformation $\epsilon^\lambda_{ij}(k)\rightarrow  \epsilon^\lambda_{ij}(k) +k_i \xi_j + k_i \xi_j   $ neither $L^a$ given in Eq.~\eqref{Lgr} or $S^a$  in Eq.~\eqref{Sgr} is separately invariant. However, the total angular momentum $J^i= L^i+ S^i$ is invariant under coordinate transformations \cite{Maggiore:2007ulw} and should be associated with an observable quantity.

\noindent The expectation value of the  angular momentum flux  in the outgoing gravitons is therefore 
\begin{eqnarray} \label{Ldot6}
\langle  \dot L^a \rangle &=&\left(-\iota\right) \frac{\kappa^2}{4} \langle 0| \left(h^{pq}T_{pq}\right)^\dagger \epsilon^{alm}
 \int \widetilde{dk} \sum_{\lambda} \epsilon^{*\lambda}_{ij}(\hat k) a_{\lambda} ^{\dagger}(k) P^{ij,rs}(\hat k) \left(k_{l} \frac{\partial}{\partial k^{m}}\right)a_{\lambda} (k)\epsilon^{\lambda}_{rs}(\hat k)\, 
 \left(h^{p^{\prime} q^{\prime}}T_{p^{\prime} q^{\prime}}\right)|0 \rangle\nonumber\\
 &=& \left(-\iota\right) \frac{\kappa^2}{4}\epsilon^{abc}\int \frac {d^3k}{(2\pi)^3 2{\omega}_k } \left(k_{b}\frac{\partial}{\partial k^{c}}  \right) \left(2\pi\right) \delta(E^{\prime}- \hbar {\omega}_{k})  {\Lambda_{ij,lm}(\hat k)}T^{*ij}(\omega^\prime)\,T^{lm}(\omega^\prime).
\end{eqnarray}
In the quadrupole approximation, $T_{ij}$ depends only on the magnitude of the graviton momenta $k=|\textbf{ k}|$ through the delta function. So the momentum operator acting on the $T_{ij}(k)$ terms will give
\begin{eqnarray}
    \cfrac{\partial}{\partial k^c} \left(T^{*ij}(k)\,T^{lm}(k)\right) = \cfrac{\partial k}{\partial k^c}\cfrac{\partial}{\partial k}  \left(T^{*ij}(k)\,T^{lm}(k)\right) = \frac{ k_c}{k}\frac{\partial}{\partial k}  \left(T^{*ij}(k)\,T^{lm}(k)\right), 
\end{eqnarray}
which contracted with the $\epsilon^{abc} k_b$ term in Eq.~\eqref{Ldot6} will give zero. The non-zero contribution will come only from the derivatives acting on the $\Lambda_{ij, lm}(\hat k)$ given in Eq.~\eqref{TTP}. This will involve the angular momentum operator acting on the unit vectors of the form
\begin{eqnarray}
    \epsilon^{abc}  k_{b}\frac{\partial}{\partial k^{c}}  \hat k^{m}=\epsilon^{abc} k_{b}\frac{\partial}{\partial k^{c}} \left(\frac{k^{m}}{k} \right)= \epsilon^{abc}  k_{b}\left(\frac{\delta^{mc}}{k} -\frac{k^m k_c} {k^2}\right) =\epsilon^{abc} \, \hat{k}_b\, \delta^{mc}.
\end{eqnarray}
We can then perform the angular integral using the expressions in Eq.~\eqref{dOmegak1}, and obtain after some algebra
\begin{equation}\label{domegaL-1}
\int d\Omega_k\, \epsilon^{abc}  k_{b}\frac{\partial}{\partial k^{c}} \Lambda_{ij,lm}(\hat k) T^{*ij}(\omega^\prime)T^{lm}({\omega^\prime})=\frac{16\pi}{15}\epsilon^{abc}  T_{bi}(\omega^\prime)T^*_{ic}(\omega^\prime).
\end{equation}
Using this in Eq.~\eqref{Ldot6}, we have the expression for the orbital angular momentum flux of gravitons from a classical source, which is given by,
\begin{eqnarray}\label{dLdt-1}
\langle   \dot L^a\rangle= \frac{16G}{15}\epsilon^{abc} \int T_{bi}(\omega^\prime)\left(-\iota\omega_{k}\right)T^{*}_{ci}(\omega^\prime)\delta(\omega^\prime-\omega_k)\, d\omega_{k}.
\end{eqnarray}
The orbital angular momentum flux carried by the gravitons is independent of $\hbar$. Similarly, the expectation  value of the flux of spin angular momentum  (Eq.~\eqref{Sgr}) in the outgoing gravitational wave is
\begin{eqnarray}\label{dSdt-1}
    \langle \dot S^a \rangle &=& \left(-2\iota\right)\frac{\kappa^2}{4}\langle 0| (h^{pq}T_{pq})^\dagger \epsilon^{alm} \int \widetilde {dk} \sum_{\lambda}  a_{\lambda}^\dagger(k) a_{\lambda} (k) \epsilon^{*\lambda}_{lj}(\hat k)\epsilon_{jm}^{\lambda}(\hat k)\left(h^{p^\prime q^\prime}T_{p^\prime q^\prime}\right)|0 \rangle\nonumber\\
    &=& \frac{32G}{15} \epsilon^{abc} \int T_{bj}({\omega}^{\prime})\left(-\iota{\omega}_{k}\right)T^*_{cj}({\omega}^{\prime})\delta({\omega}^{\prime}-{\omega}_{k}) \, d{\omega}_{k}\,.
\end{eqnarray}
The rate of spin angular momentum radiated via gravitons integrated over the direction of emission, averaged over a period of the orbit, is independent of $\hbar$. 


\section{Gravitational radiation from elliptic orbits}\label{Sec:Elliptic_Gravity}

\begin{figure}[h] 
    \centering
    \includegraphics[width=0.45\linewidth]{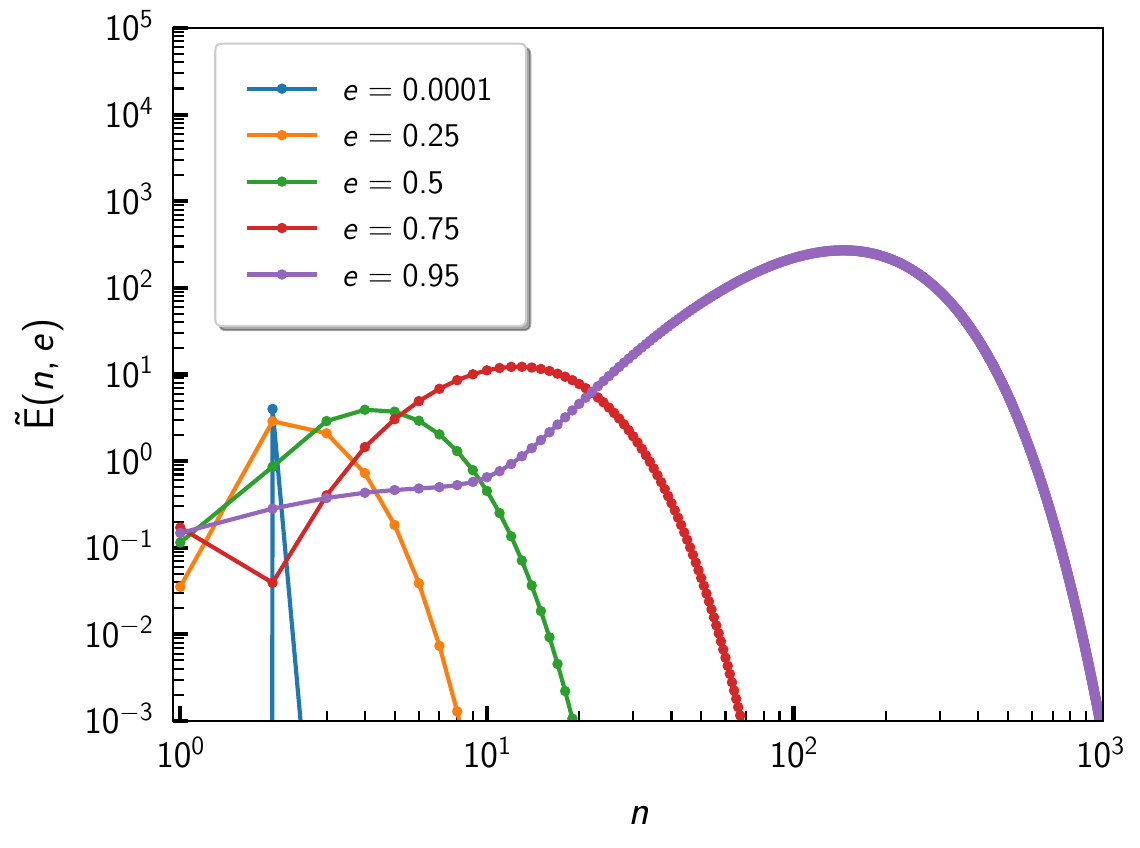}
    \includegraphics[width=0.45\linewidth]{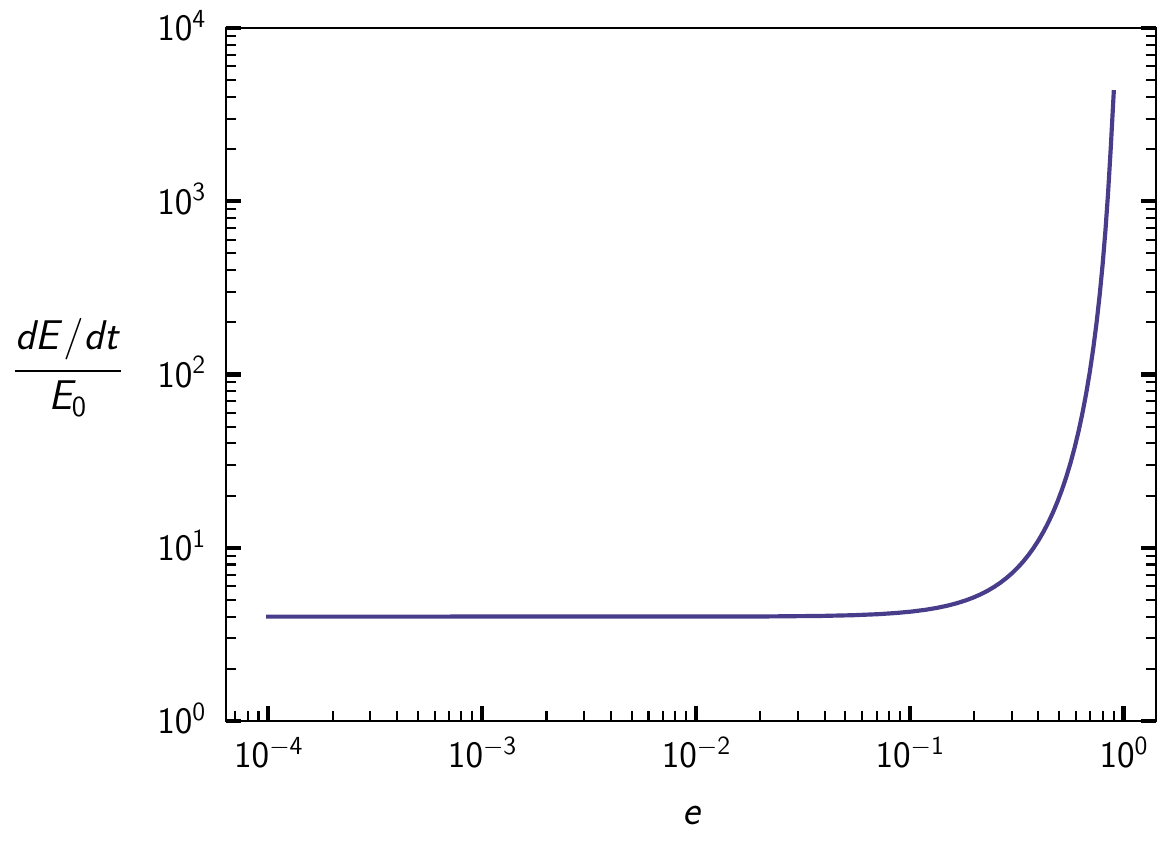}
    \caption{\textit{The left panel shows the variation of $\tilde{\mathrm{E}} (n, e)$ with the harmonic number $n$ for different values of the eccentricity $e$. At a fixed eccentricity, $\tilde{\mathrm{E}} (n, e)$ initially increases with $n$, reaches a maximum at a characteristic harmonic, and then decreases. As the eccentricity of the orbit increases, the peak shifts towards higher harmonic numbers. The right panel depicts the dependence of the radiated energy loss rate on the eccentricity. For small values of $e$, the spectrum remains almost flat; as the eccentricity increases, the energy loss rate and  eventually diverges in the limit $e\to 1$.}}
    \label{fig:loss_rate_Energy_ellp}
\end{figure}

\noindent For binaries in an elliptical orbit, the classical orbits can be described in terms of the total mass $M = m_{1} + m_{2}$ and the reduced mass $\mu = (m_1m_2)/ M$ as a one-body problem of a mass $\mu$ orbiting the center of mass with coordinates $\textbf{x}$. The Keplerian orbit can be parametrized in terms of the eccentricity of the orbit $(e<1)$, the semi-major axis $a$, and the eccentric anomaly $\xi \in (0, 2\pi)$ as follows,
\begin{eqnarray}\label{eq:elliptic-coordinate}
 x(\xi) = a\,(\cos\xi - e),\qquad y(\xi) = a \sqrt{1 - e^2}\,\sin\xi, \qquad z(\xi) = 0,\qquad \omega_0\, t = (\xi - e\sin\xi).
\end{eqnarray}
where ${\omega}_{0}=\sqrt{GM/a^3}$ is the fundamental angular frequency of the Kepler orbit. The stress tensor of this compact binary system is
\begin{equation}\label{Tijxt}
T_{ij}(\textbf{x}^{\prime}, t)={\mu}\,{\delta}^{(3)}(\textbf{x}^\prime-\textbf{x}(t))
\,\dot{x}_{i}(t) \dot{x}_{j}(t),
\end{equation}
where $\dot{x}_{i} =dx_{i}/dt$ are the components of the velocity of the reduced mass of the binary system in the Keplerian orbit described in Eq.~\eqref{eq:elliptic-coordinate} in the parametric form. The Fourier transform of the stress tensor (Eq.~\eqref{Tij}) for point particles in a periodic orbit with time period $T$, is given by 
\begin{eqnarray}
    T_{ij}(\omega^\prime) = \frac{\mu}{T}\int dt\, x_{i}(t) x_{j}(t) \,e^{\iota\omega^{\prime} t}.
\end{eqnarray}
For the Kepler orbit in Eq.~\eqref{eq:elliptic-coordinate}, we can compute the components of the stress tensor in frequency space as follows,
\begin{eqnarray}
T_{xx}(\omega^{\prime}) = \cfrac{\mu}{T} \int_{0}^{T} dt\, \dot{x}^{2}(t)\, e^{\iota \omega^{\prime} t} 
= \cfrac{-\iota\mu \omega^{\prime}}{T}\int_{0}^{T} dt\, \dot{x}(t) x(t)\, e^{\iota \omega^{\prime} t}.
\end{eqnarray}
We have used integration by parts. Using Eq.~\eqref{eq:elliptic-coordinate} we can write,
\begin{eqnarray}
\dot{x}dt = \cfrac{dx}{d\xi}\, d\xi = - a \sin \xi\, d\xi,
\end{eqnarray}
and replace $t$ in the exponent by ${\omega}_{0}^{-1}(\xi - e \sin \xi)$. We get 
\begin{eqnarray}
T_{xx}(\omega^{\prime}) = \frac{\iota\mu a^{2}\omega_{0}\,\omega^{\prime}}{2\pi } \int_{0}^{2\pi} d\xi \,\sin\xi (\cos\xi - e)\, e^{\iota(\omega^{\prime}/\omega_{0}) (\xi - e \sin\xi)},
\label{Txx2}
\end{eqnarray}
where we have replaced $T=2\pi/\omega_0$.
We can identify the integral with Bessel functions of the first kind,
\begin{eqnarray}
J_{n}(z)=\frac{1}{2\pi}\int^{2\pi}_{0} e^{i(n\xi-z\sin\xi)}d\xi,
\end{eqnarray}
by taking  $z= ne$ and $\omega^\prime= n\omega_0$ where  $n$ are non-negative integers. The Fourier modes of elliptical orbits are a sum over integer harmonics of the fundamental frequency. We denote the frequency $\omega^{\prime}$ associated with the $n$th harmonic of the fundamental as $\omega_{n}^{\prime}$, hence $\omega_{n}^{\prime} = n{\omega}_{0}$.
The components of the stress tensor Eq.~\eqref{Txx2} can then be written as linear combinations of Bessel functions as
\begin{eqnarray}\label{Txx3}
T_{xx}({\omega}_{n}^{\prime}) &=& \cfrac{\iota\mu a^{2} \omega_{n}^{\prime 2}}{2 \pi n } \int_{0}^{2\pi} d\xi \,\sin\xi (\cos\xi - e)\, e^{\iota n (\xi - e \sin\xi)} \nonumber\\
&=& \frac{\mu a^2 \omega_n^{\prime 2}}{8\pi n }\int_0^{2\pi} d\xi\,\left[(e^{2 \iota \xi}-e^{-2 \iota \xi})-2e(e^{ \iota \xi}-e^{-\iota \xi})\right]e^{\iota n (\xi - e \sin\xi)}  \nonumber\\
&=& \frac{\mu a^2 \omega_n^{\prime 2}}{4 n }\left( \left[J_{n+2}(ne)-J_{n-2}(ne) \right]-2e\left[J_{n+1}(ne) - J_{n-1}(ne)\right]\right)\nonumber\\
& =& \frac{ \mu a^{2} \omega_{n}^{\prime 2}}{n} \left[ \cfrac{1}{ e^{2}}\,\cfrac{J_{n}(ne)}{n} -\cfrac{(1 - e^{2})}{e}\,J_{n}^{\prime}(ne)  \right].
\end{eqnarray}
Where we make use of the Bessel function recurrence relations
\begin{eqnarray}\label{bessel-recurr}
J_{n-1}(z) + J_{n+1}(z) = \cfrac{2n}{z}\,J_n(z), \hspace{1cm} J_{n-1}(z) - J_{n+1}(z) = 2J^\prime_n(z),
\end{eqnarray}    
with $J^\prime_{n}(z)=dJ_n(z)/dz$.
The other non-zero components, i.e., $T_{yy}$ and $T_{xy}$ can be obtained similarly~\cite{Poddar:2021yjd, Hait:2022ukn} and are given by,
\begin{eqnarray}\label{Tyy-xy}
T_{yy}(\omega_{n}^{\prime}) &=& \cfrac{\mu}{T} \int_{0}^{T} dt\, \dot{y}^{2}(t)\,e^{\iota\omega_{n}^{\prime} t} 
 = \frac{\mu a^{2} \omega_{n}^{\prime 2}}{n}\left[\cfrac{(1 - e^{2})}{e}\, J_{n}^{\prime}(ne) - \cfrac{(1 - e^{2})}{e^{2}}\,\cfrac{J_{n}(ne)}{n} \right], \nonumber\\
T_{xy}(\omega_{n}^{\prime}) &=& \cfrac{\mu}{T}\, \int_{0}^{T}\, dt\, \dot{x}(t)\,\dot{y}(t)\, e^{\iota\omega_{n}^{\prime} t} 
= \cfrac{\iota \mu a^{2} \omega_{n}^{\prime 2}\, }{n}\left[\cfrac{\sqrt{(1 - e^{2})}}{e}\,\cfrac{J_{n}^{\prime}(ne)}{n} -\cfrac{(1-e^{2})^{3/2}}{e^{2}}\,J_{n}(ne)   \right].
\end{eqnarray}
Using the expressions for $T_{xx}, T_{yy}$, and $T_{xy}$ from Eq.~\eqref{Txx3} and Eq.~\eqref{Tyy-xy}, we compute the source term for graviton number rate in Eq.~\eqref{dNdt} and energy radiation rate in Eq.~\eqref{dEdt}, which turns out to be,
\begin{eqnarray}
    T_{ij}(\omega_n^\prime)\,T^*_{ji}(\omega_n^\prime)-\cfrac{1}{3}\,|\,T^{i}{}_{i}(\omega_n^\prime)\,|^2 = \mu^2 a^4 \omega_0^4 \,g(n, e).
\end{eqnarray}
With
\begin{eqnarray}
	g(n,e) &=& \left[\cfrac{2}{e^{4}}\left(1-e^{2}\right)^3 + \cfrac{2}{3n^{2} e^{4}}\left(3 -3e^{2} + e^{4}\right)\right]n^{2} J^{2}_{n}(ne) \nonumber + \left[\cfrac{2}{e^{2}}\left(1-e^{2}\right)^{2} + \cfrac{2}{n^{2} e^{2}}\,\left(1-e^{2}\right)\right]n^{2}J^{\prime\,2}_{n}(ne) \nonumber \\ && - \cfrac{8}{e^{3}}\left(1-e^{2}\right)\left(1-\cfrac{3}{4}\,e^{2}\right)n J_{n}(ne) J^{\prime}_{n}(ne).
\end{eqnarray}
Hence, for binaries in elliptic orbits orbit with eccentricity $e$,  the period averaged energy radiated (Eq.~\eqref{dEdt}) is
\begin{eqnarray}\label{dEdt-4}
	\cfrac{dE}{dt} &=&  \cfrac{8G}{5}  \sum_{n\, =\, 1}^\infty\left[T_{ij}(\omega_{n}^{\prime})T_{ji}^{*}(\omega_{n}^{\prime}) - \cfrac{1}{3}|T^{i}_{i}(\omega_{n}^{\prime})|^2\right] \omega_{n}^{\prime 2}
	 =  \cfrac{8G}{5}\,\mu^{2} a^{4} \omega_0^{6}\sum_{n\, =\, 1}^{\infty} n^{2} g(n,e)= E_{0}\sum_{n\,=\,1}^{\infty} n^2 g(n, e).
\end{eqnarray}
To characterize the contribution from individual modes, we introduce the quantity
\begin{eqnarray}
    \tilde{\mathrm E}(n,e) = n^2 g(n, e).
\end{eqnarray}
We illustrate the variation of $ \tilde{\mathrm E}(n,e)$ with the harmonic number for various values of eccentricity in the left panel of Fig.~\ref{fig:loss_rate_Energy_ellp}. For each eccentricity, $ \tilde{\mathrm E}(n,e)$ grows with the harmonic number, peaks at a characteristic harmonic, and then gradually decreases. The locations of the peaks shift to higher harmonics as the eccentricity increases. We show the radiated energy loss rate as a function of eccentricity in the right panel. For a nearly circular orbit, i.e., $e\ll 1$, the spectrum remains almost flat, and diverges as $e\to 1$. 

\noindent We use the Bessel sum relations derived in \cite{Peters:1963ux} and listed in Appendix~\ref{Bessel-formulas}, to sum the series in Eq.~\eqref{dEdt-4} and obtain
\begin{eqnarray}
	f(e) \equiv \sum_{n = 1}^{\infty} n^{2}g(n,e) = \cfrac{4}{(1-e^{2})^{7/2}}\left(1 + \cfrac{73}{24}\,e^{2} + \cfrac{37}{96}\,e^{4} \right).
\end{eqnarray}
Therefore, the rate of energy radiated (\ref{dEdt-4}) as gravitational waves from the binary can be written as
\begin{eqnarray}\label{PWE}
\cfrac{dE}{dt} &=& \cfrac{32G}{5}\, \mu^2 a^{4}\omega_0^{6}\,\cfrac{1}{(1 - e^{2})^{7/2}}\left(1 + \cfrac{73}{24}\,e^{2} + \cfrac{37}{96}\, e^{4} \right)\nonumber \\
&=& \cfrac{32}{5}\,G^4\,\cfrac{m^2_{1}m^2_{2}\left(m_{1} + m_{2}\right)}{a^{5} \left(1-e^2\right)^{7/2}}\left(1 + \cfrac{73}{24}\,e^{2} + \cfrac{37}{96}\, e^{4} \right).
\end{eqnarray}
This expression matches the expression for energy radiated as gravitational waves from elliptic binary orbits using the classical quadrupole formula~\cite{Peters:1963ux}. Now we compute the average rate of graviton number (Eq.~\eqref{dNdt}) radiated in binary orbit, which   is obtained by dividing each term in  the energy mode sum in  (\ref{dEdt-4}) by $n \hbar \omega_0 $ and is given by
\begin{eqnarray}\label{dNdt-4}
	\cfrac{dN}{dt}  =  \cfrac{8 G}{5 \hbar}  \sum_{n = 1}^\infty\left[T_{ij}(\omega_{n}^{\prime})T_{ji}^{*}(\omega_{n}^{\prime}) - \cfrac{1}{3}|T^{i}_{i}(\omega_{n}^{\prime})|^2\right] \omega_{n}^{\prime }
	 = \cfrac{8G}{5 \hbar} \,\mu^{2} a^{4} \omega_0^{5}\,\sum_{n = 1}^{\infty} n\, g(n,e)
    = \cfrac{N_{\rm ell}}{\hbar} \sum_{n = 1}^{\infty} n\, g(n,e)\,,
\end{eqnarray}
with $N_{\rm ell} = (8G/5)\mu^2 a^4\omega_{0}^5$. To quantify the contribution of individual modes, we introduce a quantity
\begin{eqnarray}
\tilde{N} (n, e) = n\, g(n, e)\,.
\end{eqnarray}
 We now compute the mode sum    
\begin{eqnarray}\label{nbar} 
\tilde{n} (e) \equiv \sum_{n = 1}^{\infty} n\,  g(n,e).
\end{eqnarray}
And this does not have a closed form solution as the Bessel sums of $J_n^2(ne)$ and $J_n^{\prime 2}(ne)$ with odd powers of $n$ cannot be computed in the same way as the remaining sums in Appendix~\ref{Bessel-formulas}, as has been noted in literature~\cite{Forseth:2015oua, Page:2025ncu, Chung:2024ymj}. From Eq.~\eqref{dNdt-4} and Eq.~\eqref{nbar}, we obtain the expression for the rate of graviton number radiated averaged over an orbit,
\begin{eqnarray}\label{Eq:dN_dt_grav_ellip}
 \frac{dN}{dt} = \cfrac{8G}{5 \hbar}\,\mu^{2}a^{4} \omega_0^{5}\tilde{n}(e) = \cfrac{N_{\rm ell}}{\hbar}\,\tilde{n}(e)\,. 
\end{eqnarray}
 This formula, naturally, has no counterpart in classical GR. Finally, we derive the expression for the angular momentum given in Eq.~\eqref{dLdt-1} and Eq.~\eqref{dSdt-1}, by noting that for a binary orbit in the $x-y$ plane, the angular momentum vector is along the $z$-axis. The source current for this is
\begin{eqnarray}
    \epsilon^{zxy}T_{xi}(\omega^{\prime}_{n})T^{*}_{iy}(\omega^{\prime}_{n}) = 2 T_{xy}(\omega^{\prime}_{n})\left[T_{yy}(\omega^{\prime}_{n})-T_{xx} (\omega^{\prime}_{n}) \right].
\end{eqnarray}
Using the expressions for $T_{ij}(\omega^{\prime}_{n})$, we find that 
\begin{eqnarray}
2\iota T_{xy}(\omega^{\prime}_{n}) \left[T_{yy}(\omega^{\prime}_{n})-T_{xx} (\omega^{\prime}_{n}) \right] = - 2 \mu^2 a^4 \omega_0^4 \,\sqrt{\cfrac{1-e^2}{e^2}}\, \ell(n,e),  
\end{eqnarray}
where 
\begin{eqnarray}
    \ell(n,e) = \left[\cfrac{(1-e^{2})(2-e^{2})}{n e^{3}}\right]n^{2} J^{2}_{n}(ne) + \cfrac{2(1-e^{2})}{ne}\,n^{2} J^{\prime\, 2}_{n}(ne) - \left[\cfrac{2(1-e^{2})^{2}}{n e^{2}} + \cfrac{(2-e^{2})}{n^{3} e^{2}}\right]n^{3} J_{n}(ne) J^{\prime}_{n}(ne).
\end{eqnarray}
Substituting these expressions in Eq.~\eqref{dLdt-1} and summing over all the modes, we obtain the total radiated orbital angular momentum as
\begin{eqnarray}\label{dLdt-4}
\cfrac{dL^z}{dt} =  \cfrac{32G}{15}\,  \mu^{2} a^{4} \omega_{0}^{5} \sum_{n\,=\,1}^\infty \sqrt{\cfrac{1-e^{2}}{e^{2}}}\,n\, \ell(n,e).
\end{eqnarray}
Similarly, using Eq.~\eqref{dSdt-1}, we find the expression for the spin angular momentum carried away by the emitted gravitons as
\begin{eqnarray}\label{dSdt-5}
\cfrac{dS^z}{dt}=  \cfrac{64\,G}{15}\,  \mu^2 a^4 \omega_0^5 \sum_{n\,=\,1}^\infty \sqrt{\cfrac{1-e^{2}}{e^{2}}}\,n\, \ell(n,e). 
\end{eqnarray}
The total angular momentum radiated by elliptical binaries is obtained by adding the spin and the orbital angular momenta
\begin{eqnarray}\label{dGJ}
    \cfrac{dJ}{dt}=\cfrac{dL^z}{dt}+\cfrac{dS^z}{dt}= \cfrac{32G}{5}\,  \mu^2 a^4 \omega_0^5 \sum_{n\,=\,1}^\infty \sqrt{\cfrac{1-e^{2}}{e^{2}}}\, n\, \ell(n,e)  \equiv 4N_{\rm ell}\sum_{n\,=\,1}^\infty \sqrt{\cfrac{1-e^{2}}{e^{2}}}\, n\, \ell(n,e)\,.
\end{eqnarray}
As in the case of radiated number density, we introduce the quantity $\tilde{\dot{J}}_{\rm ell} (n, e)$ to quantify the contribution of individual modes
\begin{eqnarray}
    \tilde{\mathrm {J}} (n, e) = 4\sqrt{\cfrac{1-e^{2}}{e^{2}}}\,n\,\ell(n,e).
\end{eqnarray}
The angular momentum carried by each graviton in the $n$-th harmonic is given by the ratio
\begin{eqnarray}
\mathcal{R}_{\rm ell} (n, e) \equiv \cfrac{|\tilde{\mathrm{J}}(n, e)|}{\tilde{\mathrm{N}}(n, e)} = 4\sqrt{\cfrac{1-e^{2}}{e^{2}}}\,\cfrac{|n\, \ell(n,e)|}{n\, g(n, e)}\,.
\end{eqnarray}

\begin{figure}[h]
    \centering 
    \includegraphics[width=0.495\linewidth]{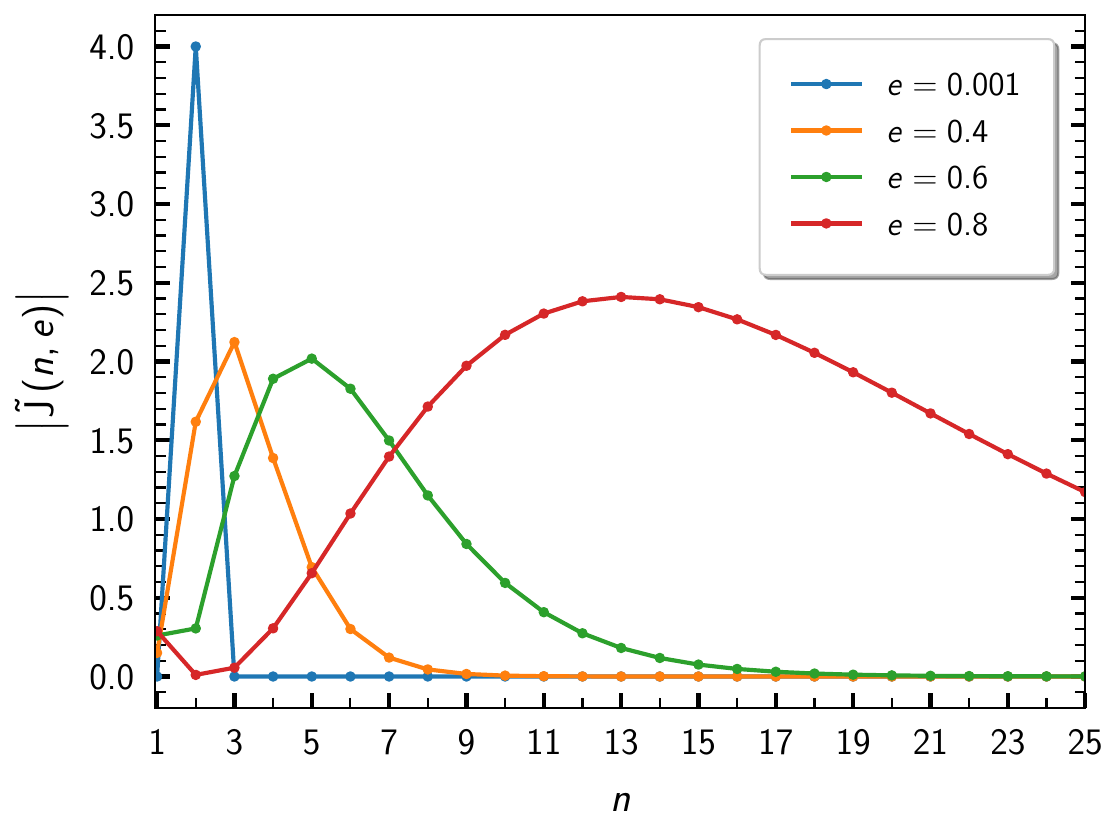}
    \includegraphics[width=0.495\linewidth]{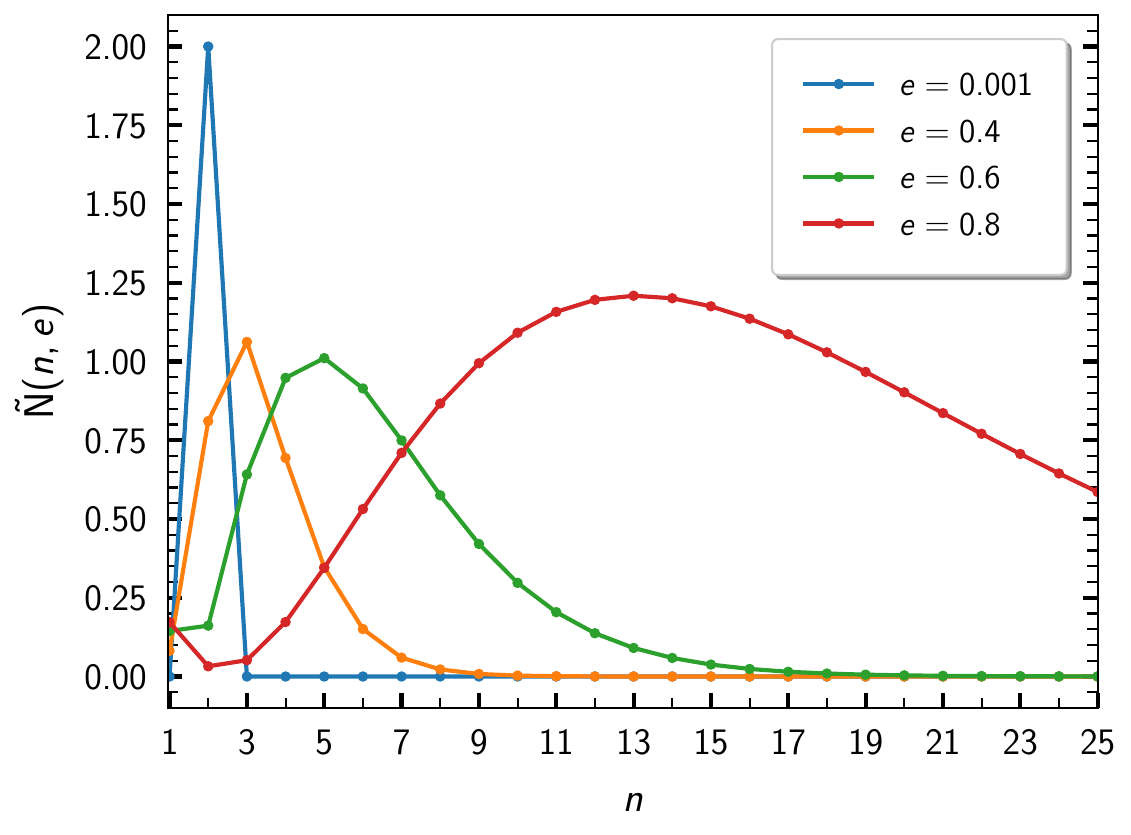}
    \includegraphics[width=0.6\linewidth]{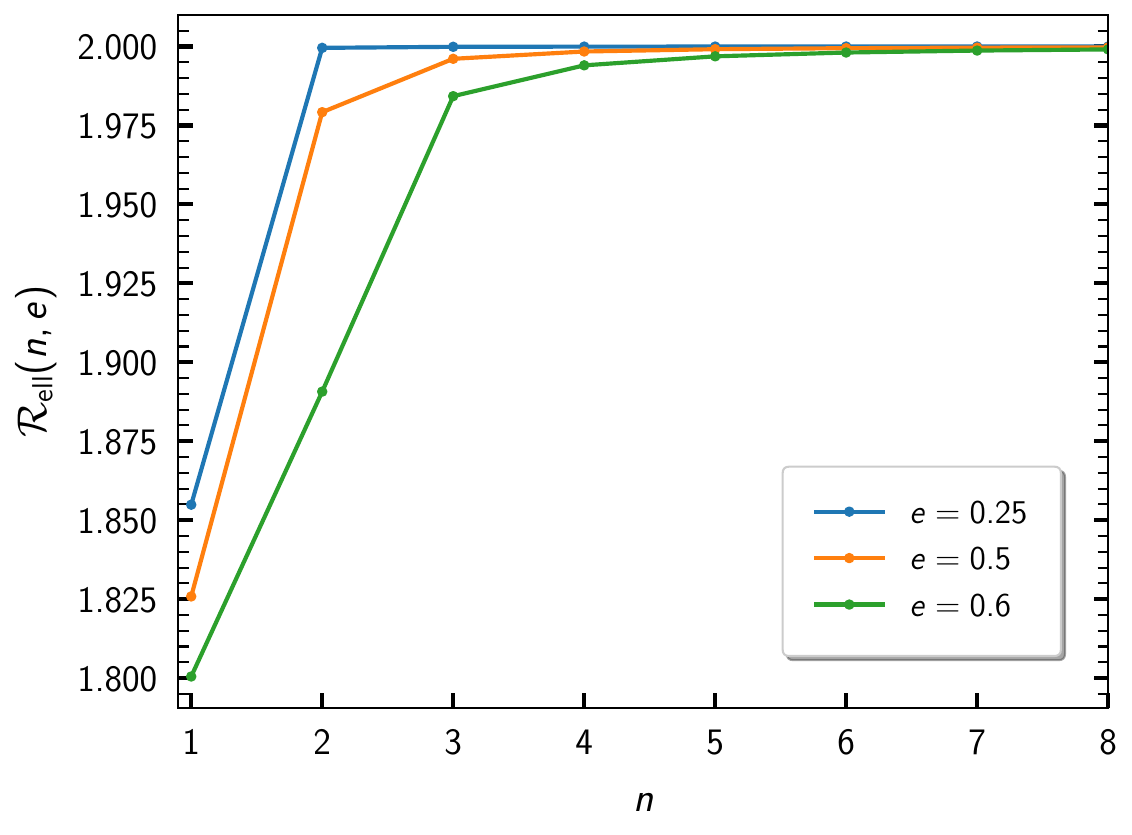}
    \caption{\it The functional dependence of $\tilde{\dot{J}}_{\rm ell} (n, e)$ and $\tilde{\dot{N}}_{\rm ell} (n, e)$ on the frequency modes is illustrated for distinct values of the orbital eccentricity. For the circular orbit ($e\to 0$), only $n=2$ mode contributes to both $\tilde{\dot{J}}_{\rm ell} (n, e)$ and $\tilde{\dot{N}}_{\rm ell} (n, e)$. Higher harmonics begin to contribute as the eccentricity increases. The bottom panel depicts the ratio $\mathcal{R}_{\rm ell}(n, e)$ as a function of the harmonic number $n$. For smaller harmonics, the ratio stays below $2$. As the harmonic number increases, it gradually approaches the asymptotic value of $2$ eventually saturates.}
    \label{fig:dotJ_dotN_ellip}
\end{figure}

Figure~\ref{fig:dotJ_dotN_ellip} highlights the functional dependence of $\tilde{\dot{J}}_{\rm ell}(n, e)$ and $\tilde{\dot{N}}_{\rm ell}(n, e)$ on the frequency modes for different values of orbital eccentricity. For the circular orbit i.e., $e \to 0$, both $\tilde{\dot{J}}_{\rm ell}(n, e)$ and $\tilde{\dot{N}}_{\rm ell}(n, e)$ receive contributions exclusively from the $n=2$ mode. With increasing eccentricity, the spectrum broadens to include higher harmonics.
The bottom panel presents the variation of the ratio $\mathcal{R}_{\rm ell} (n, e)$ with the mode number $n$ for orbital eccentricities $e=0.25$, $0.5$, $0.6$. In the circular-orbit limit ($e\to 0$), the only nonzero contribution arises from the $n=2$ mode, yielding $\mathcal{R}_{\rm ell}(n=2)=2$. As the eccentricity increases, the ratio is non-zero at higher values of $n$ and deviates from $2$.

\noindent With the aid of Bessel sum formulae written in Eq.~\eqref{B1}, Eq.~\eqref{B2} and Eq.~\eqref{B4}, we obtain
\begin{eqnarray} \label{LE}
    \tilde \ell(e)= \sum_{n=1}^\infty \sqrt{\cfrac{1-e^{2}}{e^{2}}}\,n\, \ell(n,e) = -\frac{\left(1+\cfrac{7}{8}\, e^2\right)}{(1-e^2)^2}\,.
\end{eqnarray}
From (\ref{dGJ}) and (\ref{LE}) we see angular momentum radiated by an elliptical orbit
\begin{eqnarray}\label{Eq:dJ_dt_grav_ellip}
    \cfrac{dJ}{dt} = -\cfrac{32G}{5}\,\mu^{2} a^{4}\omega_{0}^{5}\,\cfrac{\left(1+\cfrac{7}{8}\,e^2\right)}{\left(1-e^2\right)^2}\,.
\end{eqnarray}
The above expression matches the classical quadrupole formula for the angular momentum loss of a binary orbit by gravitational radiation~\cite{Peters:1964zz}. 
To obtain numerical estimates of the angular momentum loss rate and the rate of graviton number radiated, we evaluate these two quantities for the well-known Hulse-Taylor binary PSR B$1913+16$. This binary pulsar has
\begin{eqnarray}
    m_{1} \approx 1.387 M_{\odot}, \qquad m_{2} \approx 1.441 M_{\odot}, \qquad M = 2.828M_{\odot}, \qquad a = 1950100\, \mathrm{km}, \qquad e = 0.617.
\end{eqnarray}
Substituting these parameters into Eq.~\eqref{Eq:dN_dt_grav_ellip} and Eq.~\eqref{Eq:dJ_dt_grav_ellip} yields
\begin{eqnarray}
    \cfrac{dN}{dt} = 4.76\times10^{70} \,\mathrm{s}^{-1} \hbar^{-1}, \qquad \bigg|\cfrac{dJ}{dt}\bigg| = 9.47 \times 10^{70}\, \mathrm{s}^{-1} .
\end{eqnarray}

\begin{figure}[htb!]
    \centering
    \includegraphics[width=0.6\linewidth]{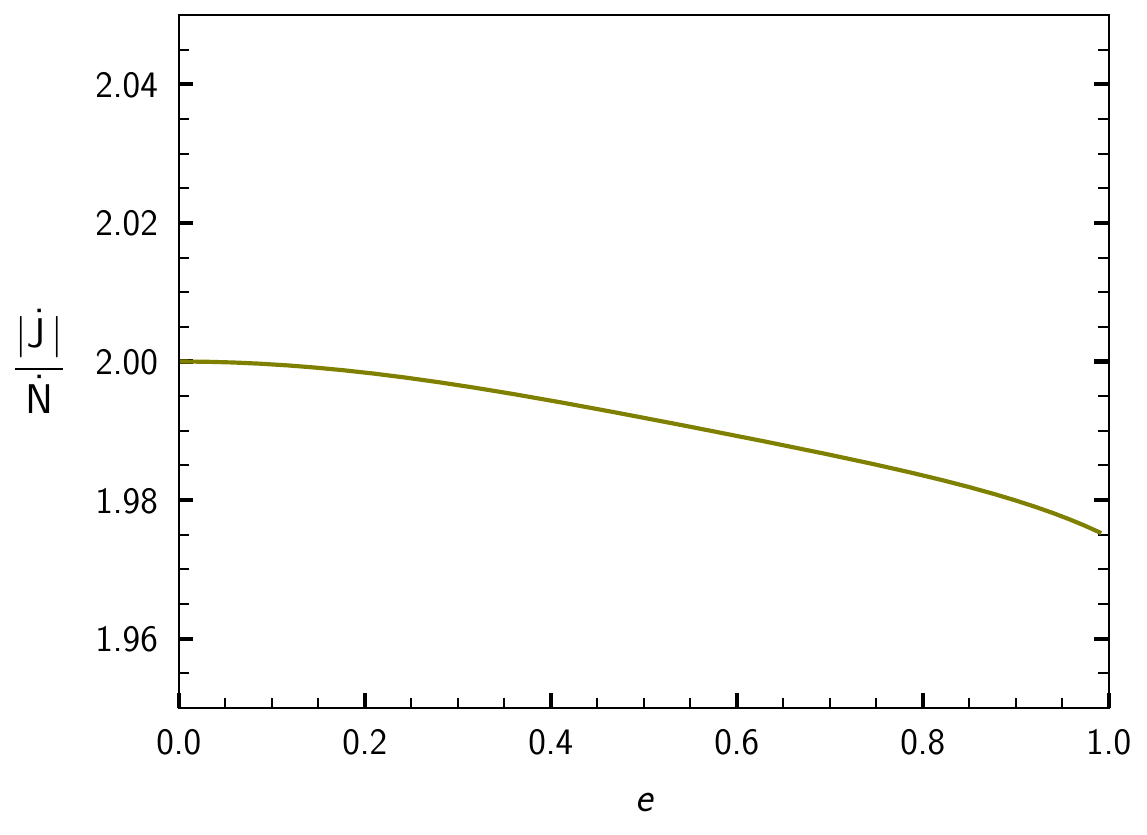}
    \caption{\it Angular momentum per emitted quantum, $\cfrac{|\dot{J}|}{\dot{N}}$ (in units of $\hbar$), plotted as a function of the eccentricity of the orbit. Irrespective of the eccentricity of the orbit, the angular momentum loss per radiated graviton from elliptic binaries stays approximately equal to $2\hbar$.}
    \label{fig:Ang_Mom_ratio}
\end{figure}

\noindent The average orbital, spin, and total angular momentum fluxes per graviton are given by
\begin{eqnarray}
    \cfrac{|dL/dt|}{dN/dt}\simeq \cfrac{2}{3}\, \hbar\, , \qquad\qquad \cfrac{|dS/dt|}{dN/dt}\simeq \cfrac{4}{3}\, \hbar\,, \qquad\qquad \cfrac{|dJ/dt|}{dN/dt}\simeq 2 \hbar \,.
\end{eqnarray}
Thus, each graviton emitted by an elliptic binary carries a total angular momentum of $2\hbar$.

\newpage

\section{Gravitational radiation from hyperbolic orbits}\label{Sec:Hyperbolic_Gravity}

\begin{figure}[htb!]
    \centering
    \includegraphics[width=0.495\linewidth]{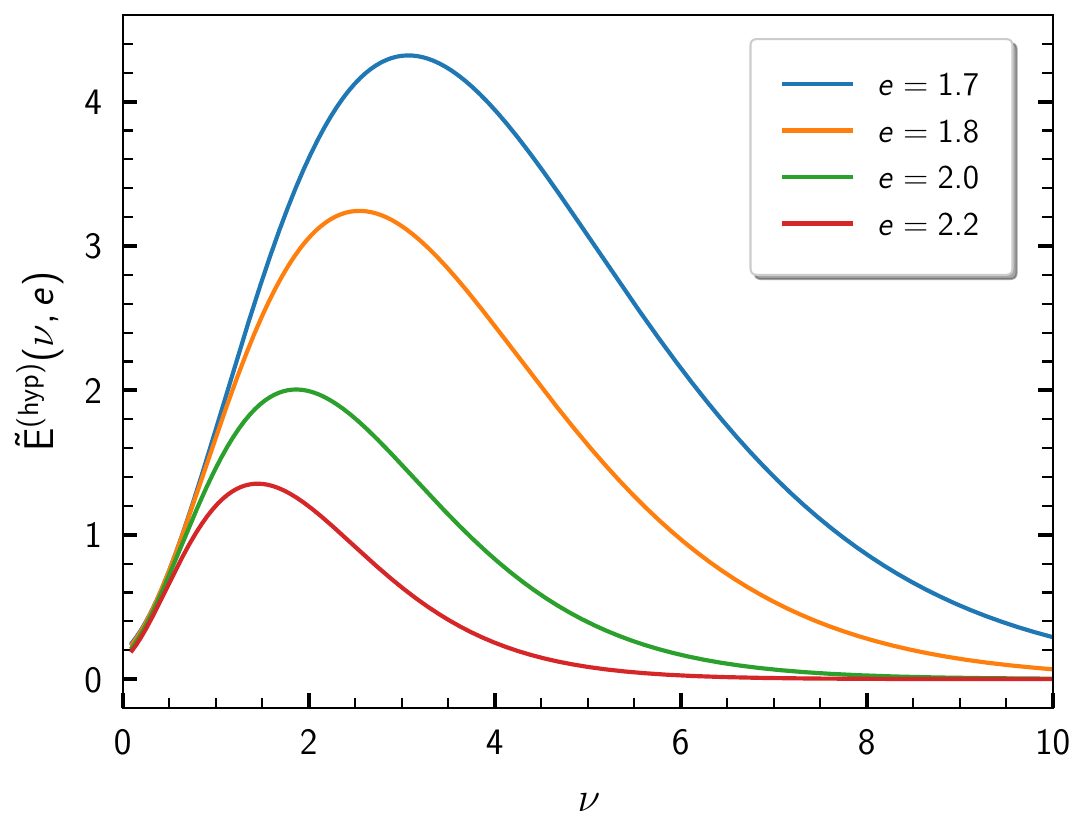}
    \includegraphics[width=0.495\linewidth]{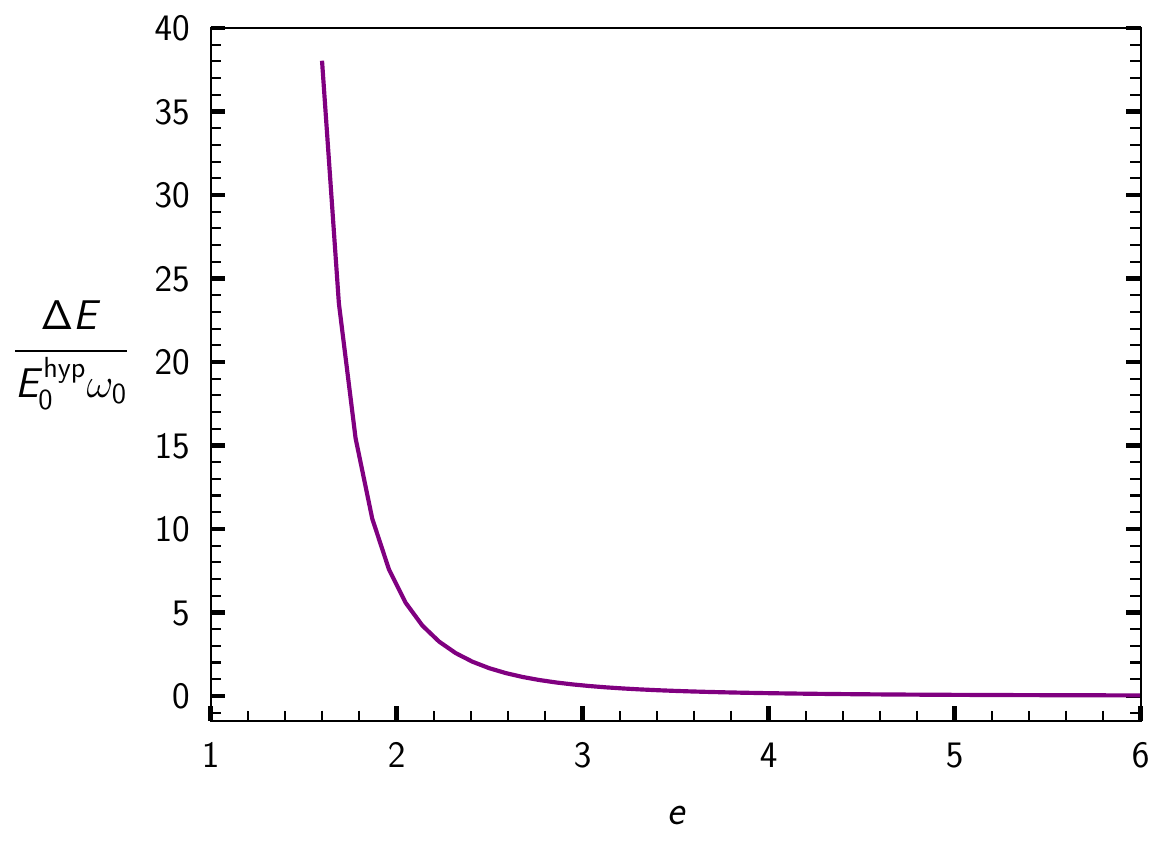}
    \caption{\it The left panel presents the energy spectrum of gravitons radiated during a hyperbolic encounter. The right panel plots the total energy radiated, integrated over all frequencies, as a function of the hyperbolic eccentricity. The energy radiated at zero frequency ($\nu = 0$) corresponds to the contribution from the gravitational wave memory signal.}
    \label{fig:Edot_nu_hyp}
\end{figure}

We consider a hyperbolic encounter between two black holes having masses $m_{1}$ and $m_{2}$, with $m_{2} > m_{1}$. The lighter black hole orbits around the heavier one. The total mass of the binary system is $M = (m_{1} + m_{2})$ having reduced mass $\mu = m_{1}m_{2}/M$.  The coordinates for the hyperbolic orbits are parametrized as
\begin{eqnarray}\label{Eq:hyp_orb_parametrization}
    x(\xi) = a\,(e - \cosh{\xi}), \qquad y(\xi) = b\sinh{\xi} = a\sqrt{e^2 - 1}\,\sinh{\xi}, \qquad z(\xi) = 0.
\end{eqnarray}
The quantities $a$ and $b$ denote the semi-major axes and the impact parameter of the orbit, respectively. The motion is confined to the $x-y$ plane, and therefore the $z$-component is set to zero. The hyperbolic anomaly $\xi \in (-\infty, \infty)$ satisfies the hyperbolic Kepler equation
\begin{eqnarray}\label{Eq:anomaly_hyp}
    (e\sinh{\xi} - \xi) = \omega_{0}\,t = \cfrac{\omega^{\prime}}{\nu} \,t.
\end{eqnarray}
The stress-tensor components for binaries in unbounded orbit can be computed as 
\begin{eqnarray}
    T_{ij}(\omega^{\prime}) = \int_{-\infty}^{\infty} dt\,e^{\iota\omega^{\prime}t}\, \dot{x}_{i}(t)\,\dot{x}_{j}(t).
\end{eqnarray}
Using the parametrization given in Eq.~\eqref{Eq:hyp_orb_parametrization} and Eq.~\eqref{Eq:anomaly_hyp}, we can express the components of the stress-tensor in terms of Hankel's function and their first derivatives as~\cite{Hait:2022ukn}
\begin{eqnarray}\label{Eq:stress-tensor}
    T_{xx}(\omega^{\prime}) &=& \left(\pi\mu a^{2}\nu\omega_{0}\right)\left[\cfrac{1}{\nu e^{2}}\,\iota H_{\iota \nu}^{(1)} (\iota e\nu) - \cfrac{(e^2 - 1)}{e}\, H_{\iota\nu}^{(1)^{\prime}}(\iota e\nu)\right]\nonumber \\
    T_{yy}(\omega^{\prime}) & = & \left(\pi \mu a^{2} \nu \omega_{0} \right)\left[\cfrac{(e^2 -1)}{\nu\,e^2}\,\iota H_{\iota \nu}^{(1)} (\iota e\nu) + \cfrac{(e^2 - 1)}{e}\, H_{\iota\nu}^{(1)^{\prime}}(\iota e\nu)\right] \nonumber \\
    \iota T_{xy}(\omega^{\prime}) &=& \left(\pi \mu a^{2} \nu \omega_{0} \right)\left[\cfrac{(e^{2} -1)^{3/2}}{e^2}\,\iota H_{\iota \nu}^{(1)} (\iota e\nu) + \cfrac{(e^2 - 1)^{1/2}}{\nu\,e}\, H_{\iota\nu}^{(1)^{\prime}}(\iota e\nu)\right].
\end{eqnarray}
The energy spectrum associated with the hyperbolic encounter is given by~\cite{Hait:2022ukn}
\begin{eqnarray}
    \cfrac{dE}{d\omega^{\prime}}\Bigg|_{\omega^{\prime}} = \cfrac{\kappa^2}{40\pi}\,\omega^{\prime2}\left(T_{ij}(\omega^{\prime})T_{ji}^{*}(\omega^{\prime}) - \cfrac{1}{3}\left|T^{i}\,_{i}(\omega^{\prime})\right|^{2}\right).
\end{eqnarray}
Using the expressions of the components of the stress-tensor given in Eq.~\eqref{Eq:stress-tensor}, we obtain 
\begin{eqnarray}
    T_{ij}(\omega^{\prime}) T_{ji}^{*} (\omega^{\prime}) - \cfrac{1}{3}\left|T^{i}\,_{i} (\omega^{\prime})\right|^{2} = \left(\pi \mu a^{2} \nu \omega_{0}\right)^2 \tilde{f}(\nu, e).
\end{eqnarray}
With
\begin{eqnarray}
    \tilde{f}(\nu, e) = \left[\alpha_{1}(\nu, e)\left(\iota H_{\iota\nu}^{(1)}(\iota e\nu)\right)^2 + \alpha_{2}(\nu, e)\left(H_{\iota\nu}^{(1)^{\prime}}(\iota e\nu)\right)^2 + \alpha_{12}(\nu, e)\,\iota H_{\iota\nu}(\iota e\nu)H_{\iota\nu}^{(1)^{\prime}}(\iota e\nu) \right].\nonumber \\
\end{eqnarray}
Here,
\begin{eqnarray}
    \alpha_{1} (\nu, e) = \cfrac{2}{e^4}\left(e^{2} -1 \right)^{3} &+& \cfrac{2e^4 -6e^2 + 6}{3\nu^2 e^4}\,, \qquad  \alpha_{2} (\nu, e) = \cfrac{2}{e^2}\left(e^2 - 1\right)^2 + \cfrac{2(e^2 -1)}{\nu^2 e^2}\,,\nonumber \\ &&  \alpha_{12}(\nu, e) = \cfrac{6(e^2 - 1)^2}{\nu e^3} - \cfrac{2(e^2 - 1)}{\nu e^3}\,.
\end{eqnarray}
The energy spectrum of gravitons radiated during a hyperbolic encounter can therefore be expressed as
\begin{eqnarray}
    \cfrac{dE}{d\omega^{\prime}}\Bigg|_{\omega^{\prime}} = \cfrac{4G}{5}\,\pi^{2} \mu^{2} a^{4} \omega_{0}^{4} \nu^{4}\tilde{f}(\nu, e) = E_{0}^{\rm hyp} \nu^4\tilde{f}(\nu, e),
\end{eqnarray}
with $E_{0}^{\rm hyp} = (4G/5)\pi^{2} \mu^{2} a^{4} \omega_{0}^{4}$. To isolate the contribution of each harmonic, we define the quantity
\begin{eqnarray}
    \tilde{\mathrm E}^{\rm (hyp)} = \nu^{4} \tilde{f}(\nu, e).
\end{eqnarray}
The total energy radiated in a single hyperbolic encounter is given by
\begin{eqnarray}
    \Delta E = E_{0}^{\rm hyp}\omega_{0}\int_{0}^{\infty} d\nu\, \nu^{4}\tilde{f}(\nu, e).
\end{eqnarray}
The energy spectrum associated with the graviton emission, and the dependence of total radiated energy normalized by $E_{0}^{\rm hyp}\omega_{0}$ on the hyperbolic eccentricity are displayed in Fig.~\ref{fig:Edot_nu_hyp}. The energy radiated at zero frequency ($\nu = 0$) corresponds to the contribution from the gravitational wave memory signal.

\begin{figure}[htb!]
    \centering
    \includegraphics[width=0.495\linewidth]{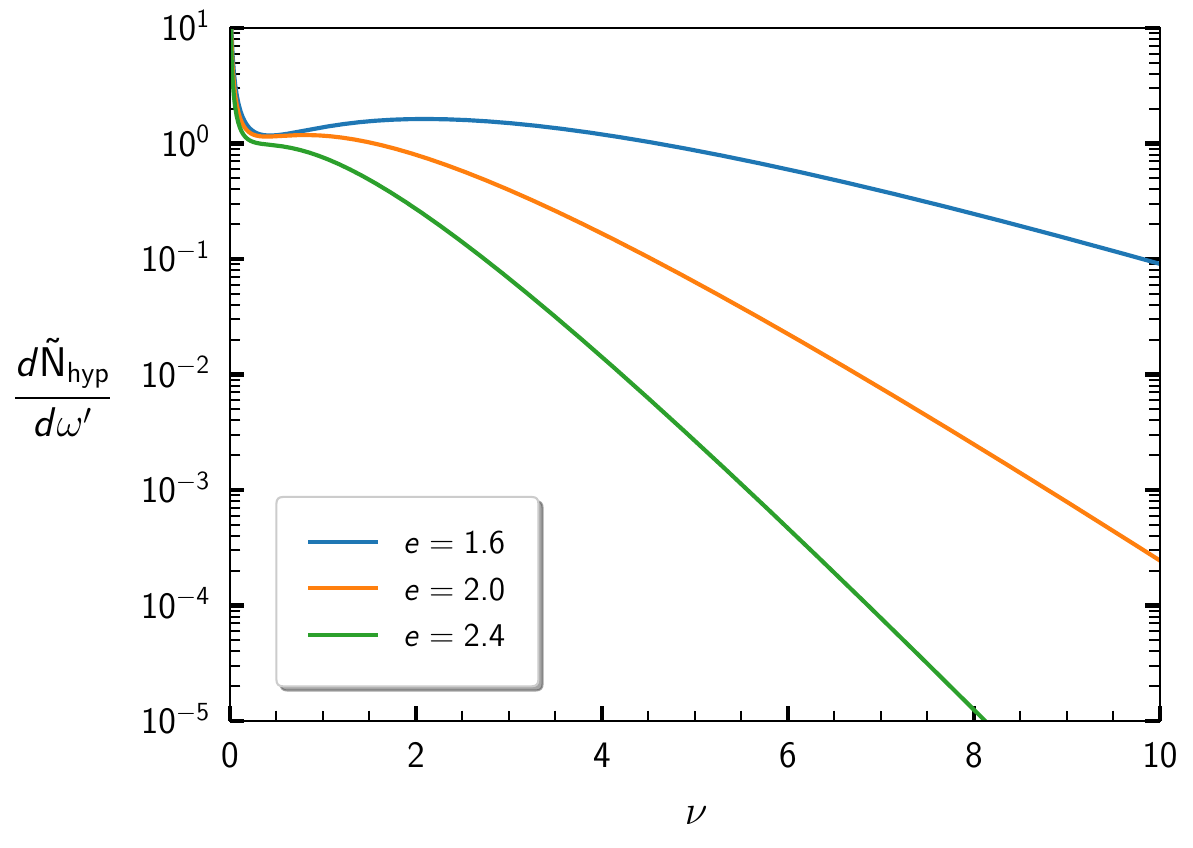}
    \includegraphics[width=0.495\linewidth]{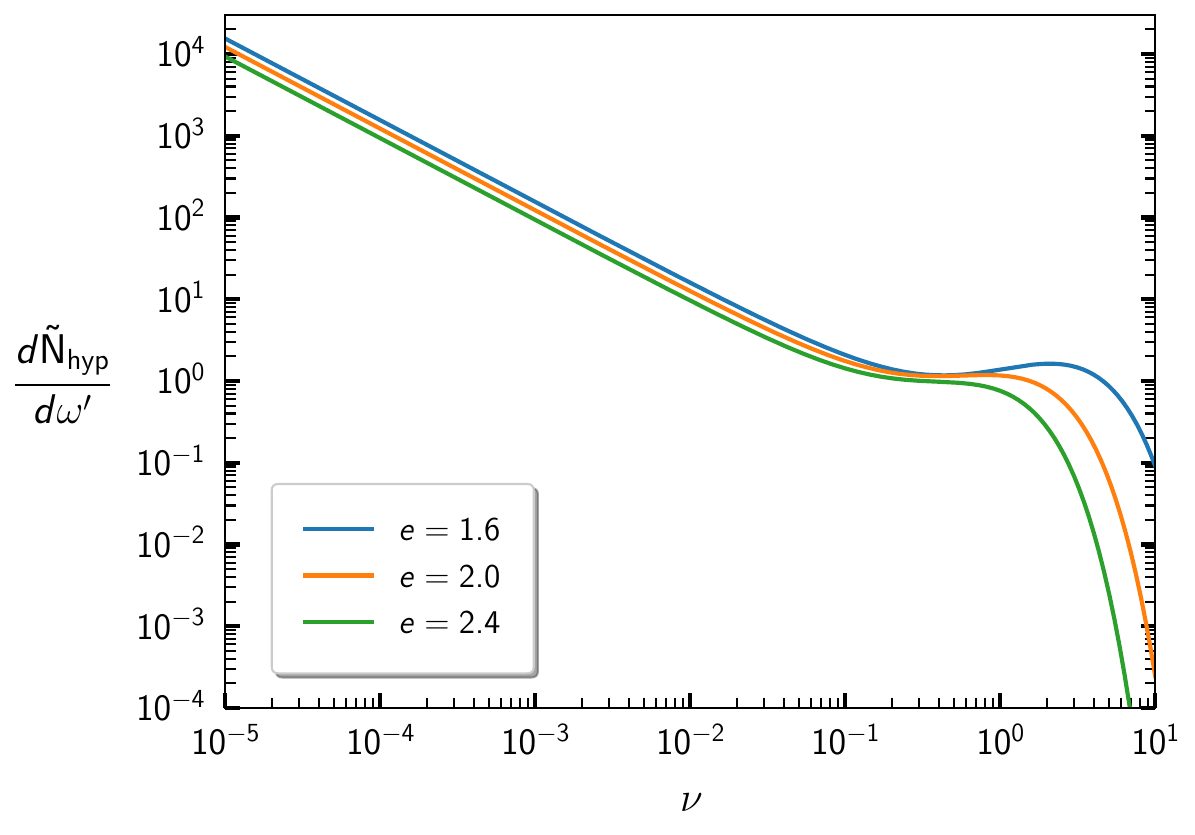}
    \caption{\it {\color{red} }The behavior of $\frac{d\tilde{N}_{\rm hyp}}{d\omega^\prime}$ as a function of $\nu$ is illustrated for hyperbolic eccentricities $e = 1.4$ (Blue), $e = 2$ (Orange), and $e = 2.4$ (Green).The left panel highlights the region $\nu \geq 1$, where the curves exhibit a rapid decline with increasing $\nu$. The right panel covers the low-$\nu$ regime. $\frac{d\tilde{N}_{\rm hyp}}{d\omega^\prime}$ diverges in the limit $\nu \to 0$.}
    \label{fig:Ndot_nu}
\end{figure}

\noindent The total number of gravitons emitted during one encounter is given by
\begin{eqnarray}\label{eq:delta_N}
    \Delta N = \int_{0}^{\infty}\cfrac{dN}{d\omega^{\prime}}\, d\omega^{\prime} 
\end{eqnarray}
With
\begin{eqnarray}\label{Eq:Number_density_loss_Hyp}
 \cfrac{dN}{d\omega^{\prime}}  \equiv \cfrac{dE/d\omega^{\prime}}{\hbar \omega^{\prime}} = \cfrac{\kappa^2}{40\pi \hbar}\,\mu^2 a^4\omega_{0}^{3}\,\pi^2\nu^3\tilde{f}(\nu, e) = \cfrac{4G}{5\hbar}\,\mu^2 a^4\omega_{0}^{3}\,\pi^2\nu^3\tilde{f}(\nu, e) \equiv \cfrac{N_{\rm hyp}}{\hbar}\,\nu^{3}\tilde{f}(\nu, e).
\end{eqnarray}
Here $N_{\rm hyp} = (4G/5)\mu^2 a^2\omega_{0}^{3}\pi^2$. Following the same approach as for the elliptical orbit, we introduce a new quantity to characterize the contribution of individual modes
\begin{eqnarray}
    \cfrac{d\tilde{N}_{\rm hyp}(\omega^{\prime}, e)}{d\omega^{\prime}} = \nu^3\tilde{f}(\nu, e).
\end{eqnarray}
Fig.~\ref{fig:Ndot_nu} illustrates the behavior of $d\tilde{N}_{\rm hyp}/d\omega^{\prime}$ for hyperbolic eccentricities $e = 1.4$ (Blue), $e = 2$ (Orange), and $e = 2.4$ (Green). The left panel focuses on the region $\nu \geq 1$, where the spectrum decreases rapidly with increasing $\nu$. The right panel covers both the low- and high-frequency regimes. As $\nu \rightarrow 0$, $d\tilde{N}_{\rm hyp}/d\omega^{\prime}$ diverges, indicating that low-frequency modes dominate the graviton number, while the contribution from high-frequency modes to $\Delta N$ is negligible.

\noindent We will estimate the rate of change of the total angular momentum due to graviton emission in a hyperbolic encounter. From the Eqs.~\eqref{dLdt-1} and ~\eqref{dSdt-1}, we can express the rate of change of angular momentum as
\begin{eqnarray}\label{Eq:Angular_Momentum_loss_Hyp}
   \cfrac{dJ}{d\omega^{\prime}} = -\cfrac{16G}{5}\,\mu^2 a^4\omega_{0}^3\,\pi^2\nu^3\, \tilde{l} (\nu, e) = -4 N_{\rm hyp}\,\nu^3\,\tilde{l}(\nu, e).
\end{eqnarray}
With
\begin{eqnarray}
    \tilde{l}(\nu, e) = \sqrt{\cfrac{e^2 -1}{e^2}}\, \left[\lambda_{1}(\nu, e)\left(\iota H_{\iota\nu}^{(1)}(\iota e\nu)\right)^2 + \lambda_{2}(\nu, e)\left(H_{\iota\nu}^{(1)^{\prime}}(\iota e\nu)\right)^2 + \lambda_{12}(\nu, e)\,\iota H_{\iota\nu}(\iota e\nu)H_{\iota\nu}^{(1)^{\prime}}(\iota e\nu) \right].\nonumber \\
\end{eqnarray}
Here,
\begin{eqnarray}
    \lambda_{1} (\nu, e) = \cfrac{(e^2 - 1)(e^2 - 2)}{\nu e^{3}}, \quad \lambda_{2} (\nu, e) = 2\,\cfrac{(e^2 - 1)}{\nu e}, \quad \lambda_{12}(\nu, e) = \left(2\,\cfrac{(e^2 - 1)^2}{e^2} + \cfrac{(e^2 - 2)}{\nu^{2} e^{2}}\right).
\end{eqnarray}
\begin{figure}[h]
    \centering
    \includegraphics[width=0.495\linewidth]{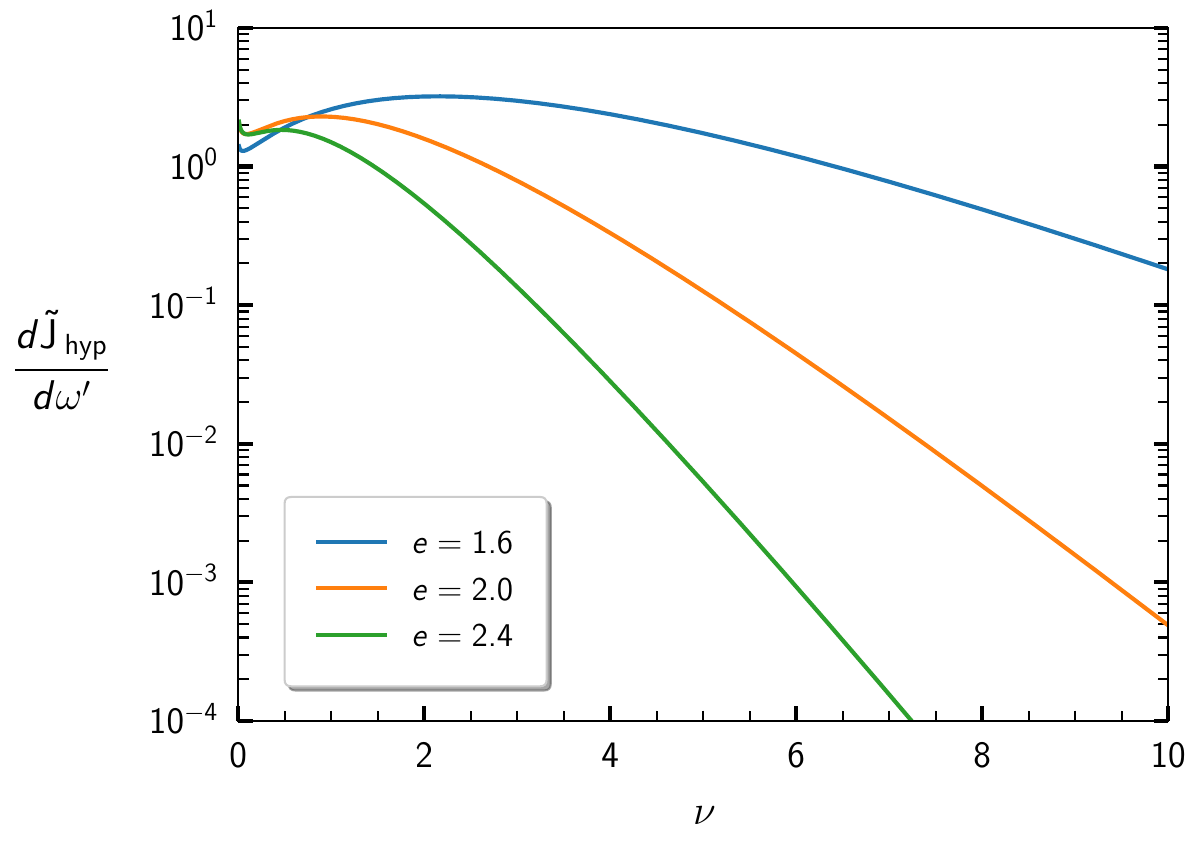}
    \includegraphics[width=0.495\linewidth]{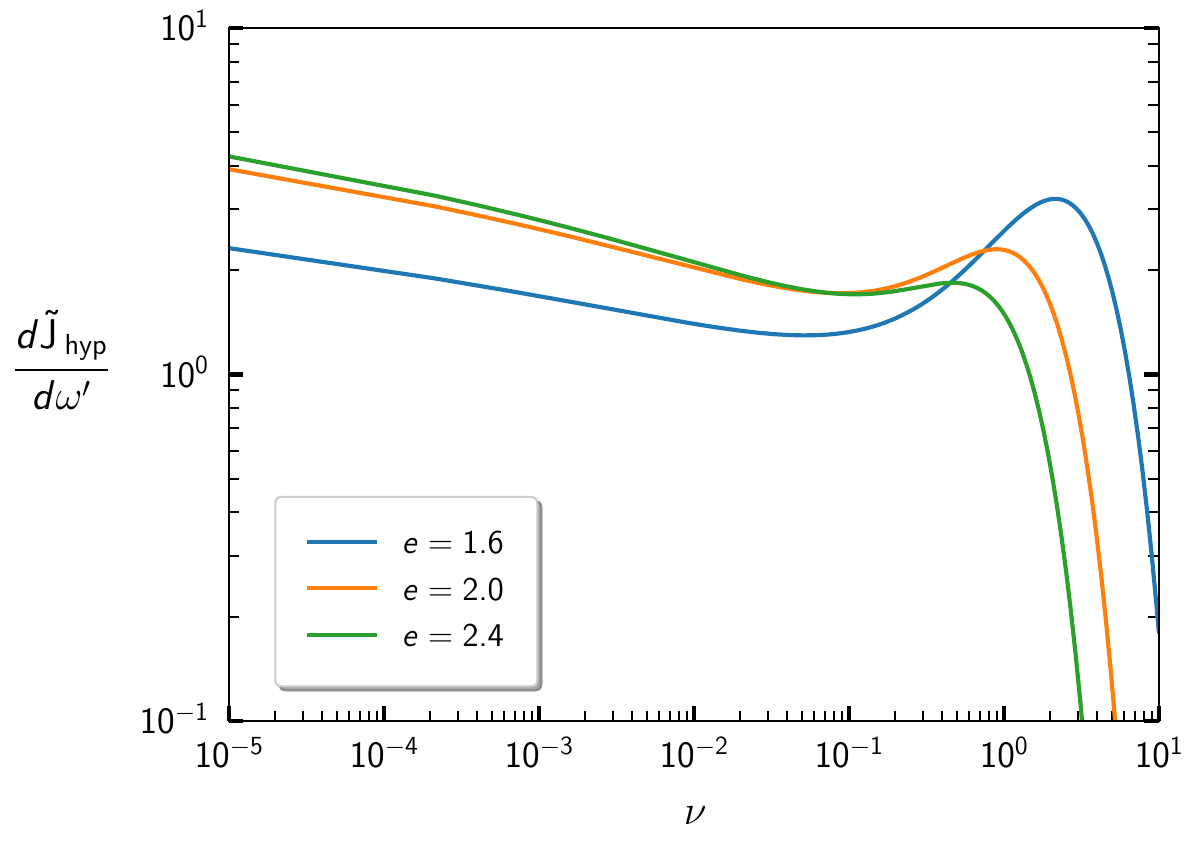}
    \caption{\it The variation of $\frac{d\tilde{J}_{\rm hyp}}{d\omega^\prime}$ with $\nu$ is shown for eccentricities $e = 1.4$ (Blue), $e = 2$ (Orange), and $e = 2.4$ (Green). The left panel covers the region $\nu \geq 1$. The curves decrease sharply with increasing $\nu$. The right panel concentrates on the low-frequency regime. In the zero frequency limit ($\nu \to 0$)$, \frac{d\tilde{J}_{\rm hyp}}{d\omega^{\prime}}$ remains finite.}
    \label{fig:Jdot_nu}
\end{figure}

\noindent To isolate the contribution of each harmonic, we define
\begin{eqnarray}
    \cfrac{d\tilde{J}_{\rm hyp}(\omega^{\prime}, e)}{d\omega^{\prime}} = 4\nu^3\,\tilde{l}(\nu, e).
\end{eqnarray}
Fig.~\ref{fig:Jdot_nu} depicts the variation of $\cfrac{d\tilde{J}_{\rm hyp}}{d\omega^{\prime}}$ with $\nu$ for different values of the eccentricity. The left panel focuses on the region $\nu \geq 1$. The curves fall off rapidly with increasing $\nu$. The right panel primarily covers the low-frequency regime. $\cfrac{d\tilde{J}_{\rm hyp}}{d\omega^{\prime}}$ remains finite as $\nu\to 0$.
\begin{figure}[htb!] 
    \centering
    \includegraphics[width=0.495\linewidth]{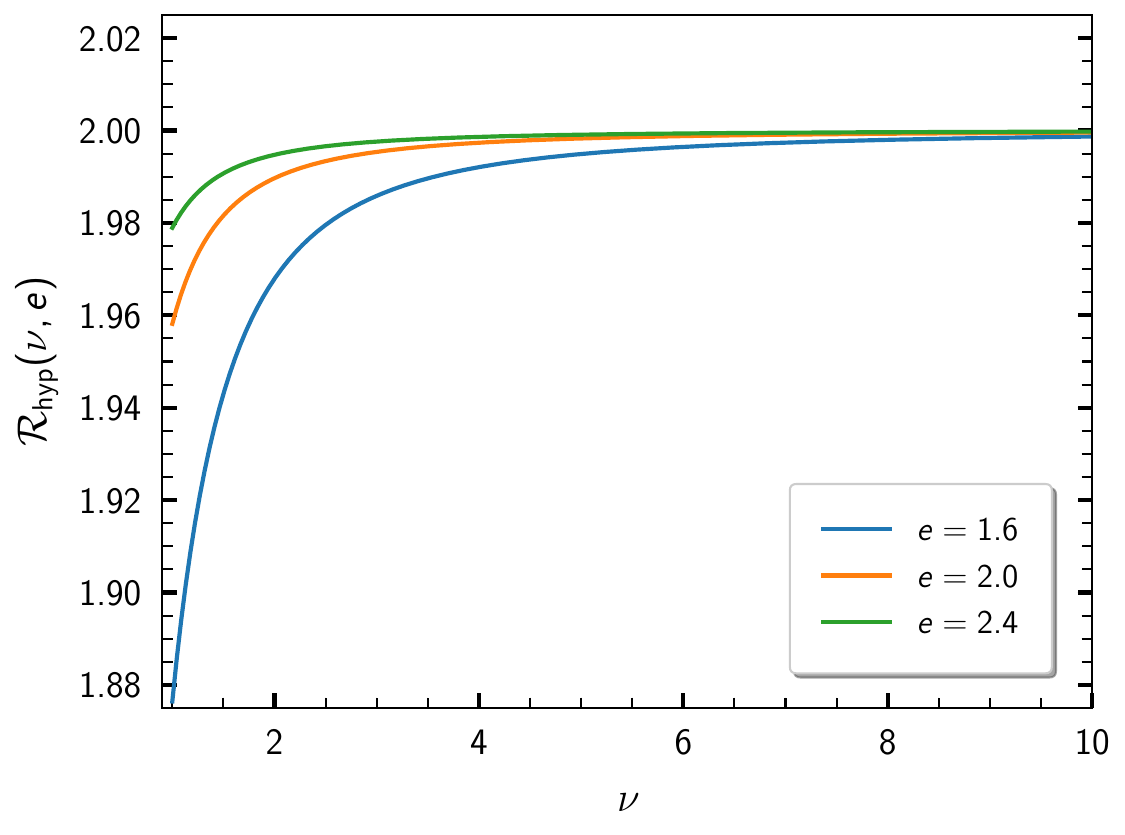}
    \includegraphics[width=0.495\linewidth]{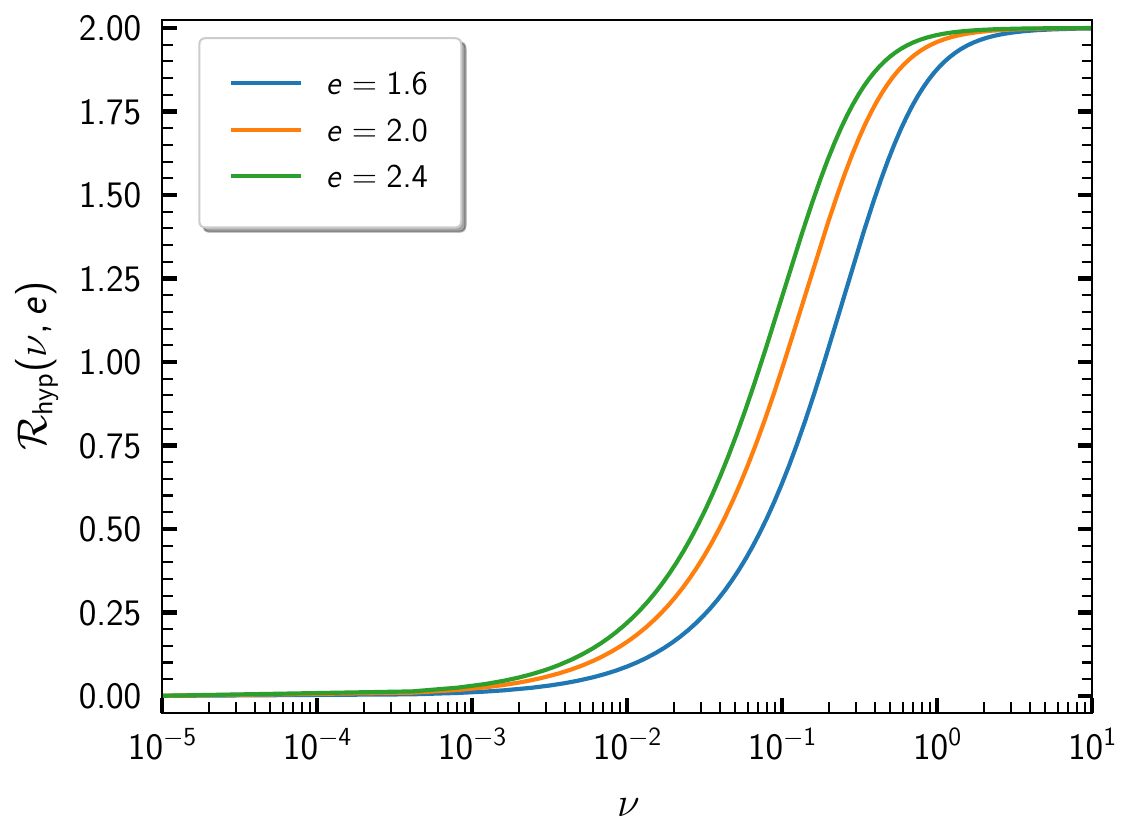}
    \caption{\it The harmonic dependence of the ratio $\mathcal{R}_{\rm hyp} (\nu, e)$ for hyperbolic encounters is depicted in the figure. The left panel represents the lower-harmonic regime, $\nu \geq 1$, whereas the right panel primarily displays the behavior in the lower-harmonic regime. The ratio starts from a value close to zero, increases with $\nu$, and eventually approaches the asymptotic value $2$.}
    \label{fig:ratio_nu_e_hyp}
\end{figure}

\begin{figure}[h]
    \centering
    \includegraphics[width=0.6\linewidth]{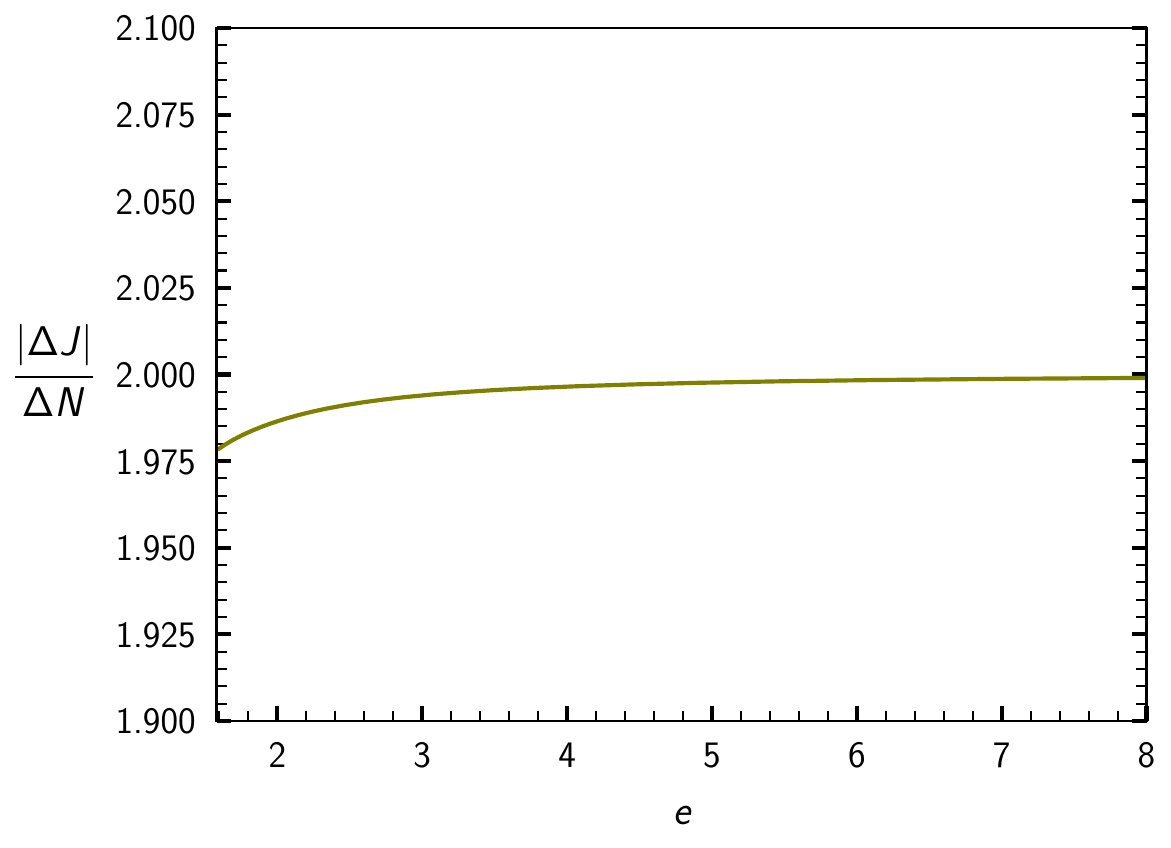}
    \caption{The ratio $|\Delta J|/\Delta N$ (in units of $\hbar$) is illustrated as a function of the eccentricity $e$. The ratio remains close to $2\hbar$ for smaller eccentricities and approaches $2\hbar$ in the limit of large hyperbolic eccentricity.}
    \label{DJDNHyp}
\end{figure}
\noindent The total radiated angular momentum in one encounter has the form
\begin{eqnarray}
    \Delta J = \int_{0}^{\infty}d\omega\,\cfrac{dJ}{d\omega}\,.
\end{eqnarray}
The radiated angular momentum per graviton in the $\nu$-th harmonic is determined by the ratio
\begin{eqnarray} \label{Rnu}
    \mathcal{R}_{\rm hyp}(\nu, e) \equiv \cfrac{d\tilde{J}_{\rm hyp}/d\omega^{\prime}}{d\tilde{N}_{\rm hyp}/d\omega^{\prime}} = 4\, \cfrac{\nu^{3}\,\tilde{l}(\nu, e)}{\nu^{3}\,\tilde{f}(\nu, e)}\,.
\end{eqnarray}
Fig.~\ref{fig:ratio_nu_e_hyp} presents the behavior of the ratio $\mathcal{R}_{\rm hyp} (\nu, e)$ as a function of $\nu$ for different values of $e$. The ratio starts from a small value, increases with $\nu$, and eventually approaches the asymptotic value $2$ at higher values of $\nu$. The right panel highlights the low-frequency regime. The ratio approaches zero as $\nu\to 0$. Fig.~\ref{DJDNHyp} illustrates the ratio $|\Delta J|/\Delta N$ (in units of $\hbar$) as a function of the eccentricity. The ratio remains close to $2\hbar$ for smaller eccentricities and approaches $2\hbar$ in the limit of large hyperbolic eccentricity.


\section{Non-gravitational emission from binary stars}\label{Sec:non_gravity_rad}

For calculating the energy radiation from binary stars in elliptical or hyperbolic orbits, we express the orbits in Fourier space as a sum over the harmonics of the fundamental frequency $\omega_0=\sqrt{G(1+\alpha)M/a^3}$. Where in the presence of a fifth force mediated by a vector exchange $\alpha = g_{V}^2 Q_{1} Q_{2}/(4\pi Gm_{1} m_{2})$, when the orbital separation is such that we have $m_{V} r \ll 1$. When the fifth force is mediated by a scalar exchange $\alpha=- g_s^2 N_1 N_2/(4 \pi G m_1 m_2)$ when $m_s r \ll 1$.  the For compact binaries separated by distance $a\simeq 1\, {\rm A.U}$ and masses $m_i\simeq 1.4\, M_\odot$ the fundamental frequency turns out to be $\omega_0 \simeq 10^{-20} {\rm eV}$. This is in the mass range of ultralight dark matter~\cite{Hu:2000ke, Hui:2016ltb, Schive:2025bcm}. In this section, we compute the energy and angular momentum radiated by scalar dark matter like axions~\cite{KumarPoddar:2019jxe} and dark photons~\cite{KumarPoddar:2019jxe}. We then compute the angular momentum per quantum of radiation for the scalar and vector cases, as we did for gravitational radiation, to examine whether the spin of radiated fields can be determined from observations.

\subsection{Radiation of scalars from binary stars}

After deriving the angular momentum per quantum for graviton radiation from eccentric binaries, we now turn to estimate the angular momentum loss per quantum due to scalar emission. Massless scalar fields with couplings to baryons can appear in scalar-tensor theories of gravity~\cite{Brans:1961sx, Dicke:1961gz, Weinberg:1972kfs}, and as dilatons associated with spontaneously broken conformal symmetry~\cite{Fujii:1971vv}. We consider a scalar field that couples to the baryon through the coupling 
\begin{eqnarray}\label{scalar-psi}
\mathcal{L}_{\rm{int}} = g_{s} \phi\, \bar{\psi} \psi,
\end{eqnarray}
with $g_s$ being the dimensionless coupling. For a macroscopic baryon source, the above interaction can be recast in terms of the baryon number density $n(x)$ as 
\begin{eqnarray}\label{scalar-n}
    \mathcal{L}_{\rm{int}} = g_{s} \phi\, n(x).
\end{eqnarray}
The baryon number density corresponding to the binary stars (indexed by $a = 1,2$) is represented by
\begin{eqnarray}\label{number-density-x}
    n(x^\prime) = \sum _{a\, =\, 1, 2}  N_{a}  \delta^{(3)} (x^\prime - x_{a}(t)).
\end{eqnarray}
Here $N_{a}(\mathcal{O}(10^{57}))$ denotes the total number of baryons that comprise the neutron star and $x_{a}(t),\,(a=1,2)$ characterizes the Keplerian trajectory of the $a$-th star in the binary system. The rate of scalar particle emission from the classical source $n (x)$ is given by
\begin{eqnarray}\label{emission-rate}
    d\Gamma = g_{s}^{2} |n(\omega^{\prime},\mathbf{k}^\prime)|^{2}(2\pi) \delta(\omega - \omega^{\prime})\,\cfrac{d^{3}p}{(2\pi)^{3}2\omega}\,.
\end{eqnarray}
The rate of energy loss due to the scalar radiation can be written as,
\begin{eqnarray}\label{eq:rate-of-energy-loss-scalar}
    \cfrac{dE_{s}}{dt} = \int \omega\, d\Gamma &=&  g_{s}^{2}\int  |n(\omega^{\prime},\mathbf{k}^\prime))|^{2} \omega (2\pi) \delta(\omega - \omega^{\prime})\,\cfrac{p^{2} dp d\Omega}{(2\pi)^{3} 2\omega} \nonumber \\
     &=& \cfrac{g_{s}^{2}}{8\pi^{2}}\int  |n(\omega^{\prime},\mathbf{k}^\prime))|^{2} \omega^{2} \sqrt{1 - \cfrac{m_{\phi}^{2}}{\omega^{2}}}\,  
 \, \delta(\omega - \omega^{\prime})  d\Omega d\omega.
\end{eqnarray}
Where we have used the mass-shell relations for the emitted scalars $\omega^2=p^2+m_\phi^2$ with $p=|\mathbf{p}|$. $n(\omega^\prime,\mathbf{k}^\prime)$ refers to the baryon number density of the source (Eq.~\eqref{number-density-x}) in Fourier space and is given by
\begin{eqnarray}\label{dipole-number-1}
    n(\omega^{\prime},\mathbf{k}^\prime)) = \cfrac{1}{T} \int_{0}^{T}dt \int d^{3}\mathbf{x}^\prime\, e^{-\iota \mathbf{k}^{\prime} \cdot \mathbf{x}^\prime} e^{\iota \omega^{\prime} t} \sum_{a\,=\, 1, 2}  N_{a} \delta^{(3)}(\mathbf{x}^\prime - \mathbf{x}_{a}(t))
    = \sum_{a\, =\, 1, 2} N_{a} \cfrac{1}{T} \int_{0}^{T} dt \, e^{-\iota \mathbf{k}^{\prime}\cdot\mathbf{x}_{a}}  e^{\iota \omega^{\prime} t}.
\end{eqnarray}
For non-relativistic motion of bounded sources $\mathbf{k}^{\prime}\cdot\mathbf{x}_{a} \ll \omega^{\prime}\, t$, therefore we can Taylor expand $e^{\iota \mathbf{k}^{\prime} \cdot\mathbf{x}_{a}}$ and in the dipole approximation we retain up to  the terms linear in $\mathbf{k}^{\prime} \cdot \mathbf{x}_{a}$ to obtain,
\begin{eqnarray}\label{nomega-1}
    n(\omega^{\prime},\mathbf{k}^\prime))
    = \cfrac{1}{T}  \left[ N_1 \int_{0}^{T} dt\left( 1 - \iota \mathbf{k}^{\prime} \cdot \mathbf{x}_{1} \right) e^{\iota \omega^{\prime} t } + N_{2} \int_{0}^{T} dt\left( 1 - \iota \mathbf{k}^{\prime} \cdot \mathbf{x}_{2} \right) e^{\iota \omega^{\prime} t } \right].
\end{eqnarray}
The stellar coordinates, $\mathbf{x}_1$ and $\mathbf{x}_2$, can be written in terms of the center of mass coordinates of the reduced mass orbit $\mathbf{x}$ as 
\begin{eqnarray}\label{eq:com-coordinates}
    \mathbf{x}_{1} = \cfrac{m_{2}\,\mathbf{x}}{m_{1}+m_{2}} = \cfrac{\mu}{m_{1}}\,\mathbf{x}, \qquad\text{and}\qquad \mathbf{x}_{2} = - \cfrac{m_{1}\,\mathbf{x}}{m_{1}+m_{2}} = - \cfrac{\mu}{m_{2}}\,\mathbf{x}.
\end{eqnarray}
In the center-of-mass (COM) coordinates we can write Eq.~\eqref{nomega-1} as
\begin{eqnarray}\label{nomega-2}
    n(\omega^{\prime},\mathbf{k}^\prime))  = \cfrac{1}{T} (N_{1} + N_{2})\int_{0}^{T} dt\,e^{\iota \omega^{\prime} t } + \iota \mu \left( \cfrac{N_{1}}{m_{1}} - \cfrac{N_{2}}{m_{2}} \right)  \cfrac{1}{T} \int_{0}^{T} dt\,\mathbf{k}^{\prime} \cdot \mathbf{x}\,e^{\iota \omega^{\prime} t }.
\end{eqnarray}
If the particle number $(N_1+N_2)$ in the compact stars is conserved, then the first monopole term gives a delta function in $\omega^{\prime}$, which will give zero contribution to the scalar number in Eq.~\eqref{emission-rate} and radiated energy in Eq.~\eqref{eq:rate-of-energy-loss-scalar}. Here, $\mathbf{k}^{\prime}$, is oriented along the direction of emission of the primary waveform, $\hat{n}^\prime = (\sin{\theta^{\prime}} \cos{\phi^{\prime}}, \sin{\theta^{\prime}} \sin{\phi^{\prime}}, \cos{\theta^{\prime}} )$. Writing the stellar coordinates in the center of mass frame, using Eq.~\eqref{eq:com-coordinates}, and incorporating the rotation of the binary plane,
\begin{eqnarray}\label{eq:coordinate-rotation}
x^{\prime\,i} = \mathcal{R}^{i}{}_{j} x^j, \quad \text{i.e.} \quad
    \begin{pmatrix}
    x^{\prime}\\ y^{\prime} \\ z^{\prime} \\
    \end{pmatrix}
     = \begin{pmatrix}
     1\quad & 0\quad & 0 \\
     0\quad & \cos{i}\quad & - \sin{i} \\
     0\quad & \sin{i}\quad & \cos{i} \\
     \end{pmatrix}
     \begin{pmatrix}
         x \\
         y \\
         z \\
     \end{pmatrix}
     = \begin{pmatrix}
         x \\ 
         y \cos{i} \\
         y \sin{i} \\
     \end{pmatrix}\,,
\end{eqnarray}

\noindent as described in Eq.~\eqref{eq:coordinate-rotation}, we can express the number density in Fourier space as
\begin{eqnarray}\label{eq:number-density-angular}
    n(\omega^{\prime}) = \left( \cfrac{N_{1}}{m_{1}} - \cfrac{N_{2}}{m_{2}} \right) \left( - \iota \mu k \right)\left[ \mathcal{F}_{1} (\theta^\prime, \phi^\prime) x(\omega^\prime) +  \mathcal{F}_{2} (\theta^\prime, \phi^\prime) y(\omega^\prime) \right], 
\end{eqnarray}
with
\begin{eqnarray}\label{eq:x-y-coeff-angular}
    \mathcal{F}_{1} (\theta^\prime, \phi^\prime) =  \sin{\theta^{\prime}}\cos{\phi^{\prime}} \qquad \text{and} \qquad \mathcal{F}_{2} (\theta^\prime, \phi^\prime) = \sin{\theta^{\prime}}\sin{\phi^{\prime}}\cos{i} + \cos{\theta^{\prime}}\sin{i}.
\end{eqnarray}
Substituting for the Fourier components of the coordinates $x(\omega^\prime)$, $y(\omega^\prime)$ in terms of Bessel functions, as given in Eq.~\eqref{eq:fourier-coord-x-y}, and taking the modulus of Eq.~\eqref{eq:number-density-angular} yields
\begin{eqnarray}\label{eq:number-density-squared-Bessel}
    |n(\omega^{\prime})|^{2} &=& \left(\cfrac{N_{1}}{m_{1}} - \cfrac{N_{2}}{m_{2}} \right)^{2}\mu^{2}|k|^{2} \Bigg[\mathcal{F}_1 (\theta^\prime, \phi^\prime)^2\, \cfrac{a^{2}}{n^{2}}\, J^{\prime 2}_{n} (ne) 
    + \mathcal{F}_2 (\theta^\prime, \phi^\prime)^2 \, \cfrac{a^{2}(1-e^{2})}{n^{2}e^{2}} \, J^{2}_{n}(ne) \Bigg] \nonumber\\
     &=&  \left(\cfrac{N_{1}}{m_{1}} - \cfrac{N_{2}}{m_{2}} \right)^{2}\mu^{2} a^{2} \omega_{0}^{2} \left(1 - \cfrac{m_{\phi}^2}{n^2\omega_{0}^2}\right) \Bigg [\mathcal{F}_{1} (\theta^\prime, \phi^\prime)^2  J^{\prime 2}_{n} (ne) + \mathcal{F}_2 (\theta^\prime, \phi^\prime)^{2} \cfrac{(1-e^{2})}{e^{2}}\,J^{2}_{n}(ne) \Bigg],
\end{eqnarray}
where we have used $|k|^2 = \omega^{\prime 2}  = (n^{2}\omega_{0}^{2} - m_{\phi}^{2})$. We obtain the flux of radiated energy owing to the scalar emission by inserting the above expression in Eq.~\eqref{eq:rate-of-energy-loss-scalar}, and it is given by
\begin{eqnarray}\label{Eq:energy-flux-theta-phi-e}
    \cfrac{dE_{s}}{dt d\Omega} 
    = \cfrac{g_{s}^{2}}{8\pi^{2}} \left( \cfrac{N_{1}}{m_{1}} - \cfrac{N_{2}}{m_{2}} \right)^{2}\mu^{2} a^{2} \omega_{0}^{4} 
    \sum_{n\,\geq\, n_{0S}}^{\infty}n^{2} \left(1 - \cfrac{n_{0S}^{2}}{n^2} \right)^{3/2} \left[\mathcal{F}_{1} (\theta^\prime, \phi^\prime)^2  J^{\prime 2}_{n}(ne) + \mathcal{F}_2 (\theta^\prime, \phi^\prime)^2 \cfrac{(1-e^{2})}{e^{2}}\,J^{2}_{n}(ne)  \right],
\end{eqnarray}
where $n_{0S} = m_{\phi}/\omega_{0}$ is the harmonic of the fundamental frequency $\omega_0$, which, in the presence of a scalar mediated fifth force, is given by $\omega_0=\sqrt{G(1+\alpha)M/a^3}$, with $\alpha=- g_s^2 N_1 N_2/(4 \pi G m_1 m_2)$ when the separation between the masses satisfies $m_\phi r \ll 1$.  Performing the integration of Eq.~\eqref{Eq:energy-flux-theta-phi-e} over the phase-space leads to the expression of the rate of energy radiated due to scalar emission
\begin{eqnarray}\label{dES}
    \cfrac{dE_{s}}{dt} = \cfrac{g_{s}^2}{6\pi} \left(\cfrac{N_{1}}{m_{1}}-\cfrac{N_{2}}{m_{2}}\right)^2 \mu^2 a^2 \omega_{0}^4 \sum_{n\,\geq\, n_{0S}}^{\infty} n^{2} \left(1 - \cfrac{n_{0S}^{2}}{n^2} \right)^{3/2} \left[ J^{\prime 2}_{n}(ne) + \cfrac{(1-e^{2})}{e^{2}}\,J^{2}_{n}(ne) \right].
\end{eqnarray}
For a massless scalar ($m_{\phi} = 0$) emission, the above equation reduces to
\begin{eqnarray}\label{Eq:energy_loss_scalar_1}
    \cfrac{dE_{s}}{dt}\Bigg|_{m_{\phi}=0} &=& \cfrac{g_{s}^{2}}{6\pi} \left( \cfrac{N_{1}}{m_{1}} - \cfrac{N_{2}}{m_{2}} \right)^{2} \mu^{2} a^{2} \omega_{0}^{4} 
    \sum_{n\, =\, 1}^{\infty}n^{2} \left[J^{\prime 2}_{n}(ne) + \cfrac{(1-e^{2})}{e^{2}}\,J^{2}_{n}(ne) \right] \nonumber \\ 
    &=& \cfrac{g_{s}^{2}}{12\pi} \left( \cfrac{N_{1}}{m_{1}} - \cfrac{N_{2}}{m_{2}} \right)^{2}\cfrac{\mu^{2} a^{2} \omega_{0}^{4}}{\left(1-e^2\right)^{5/2}}\,\left(1+\cfrac{e^2}{2}\right)\,.
\end{eqnarray}
Accordingly, the number of radiated massless scalar quanta is obtained by dividing each term in the energy mode sum of Eq.~\eqref{Eq:energy_loss_scalar_1} by $n \hbar \omega_0$, yielding
\begin{eqnarray}\label{Eq:Rate_number_loss_Scalar}
    \cfrac{dN_{s}}{dt}\Bigg|_{m_{\phi}=0} 
    =  \cfrac{g_{s}^2}{6\pi\hbar} \left(\cfrac{N_{1}}{m_{1}}-\cfrac{N_{2}}{m_{2}}\right)^2 \mu^2 a^2 \omega_{0}^3 \sum_{n\,=\,1}^{\infty} n \left[J^{\prime 2}_{n}(ne) + \cfrac{(1-e^{2})}{e^{2}}\,J^{2}_{n}(ne) \right]
\end{eqnarray}
The angular momentum loss rate for scalar emission can be expressed as
\begin{eqnarray}\label{Eq:ang_mom_scalar}
    \cfrac{dJ^{i}_{s}}{dt} &=& g_{s}^{2} \left(\cfrac{N_{1}}{m_{1}} - \cfrac{N_{2}}{m_{2}}\right)^2 \mu^2 \omega_{0}^2 \left(1- \cfrac{n_{0S}^2}{n^2}\right)\int \epsilon^{ijk} x_{j} p_{k} x_{l}(\omega^{\prime}) x_{l}^{*}(\omega^{\prime}) (2\pi) \delta (\omega -\omega^{\prime})\,\cfrac{d^{3}\omega^{\prime}}{(2\pi)^3 2\omega^{\prime}}\nonumber \\
    &=& \cfrac{g_{s}^2}{6\pi} \left(\cfrac{N_{1}}{m_{1}} - \cfrac{N_{2}}{m_{2}}\right)^2 \mu^2 \omega_{0}^2 \left(1- \cfrac{n_{0S}^2}{n^2}\right)^{3/2}\int \epsilon^{ijk} x_{j} \left(-\iota\cfrac{\partial}{\partial x_{k}}\right)x_{l}(\omega^{\prime}) x_{l}^{*}(\omega^{\prime}) \omega^{\prime} \delta(\omega-\omega^{\prime}) d\omega^{\prime}\nonumber \\
    & =& (-\iota)\cfrac{g_{s}^2}{6\pi} \left(\cfrac{N_{1}}{m_{1}} - \cfrac{N_{2}}{m_{2}}\right)^2 \mu^2 \omega_{0}^2 \left(1- \cfrac{n_{0S}^2}{n^2}\right)^{3/2}\int \left(x(\omega^{\prime}) y^{*}(\omega^{\prime}) - x^{*}(\omega^{\prime}) y(\omega^{\prime})\right) \omega^{\prime} \delta(\omega - \omega^{\prime}) d\omega^{\prime}.
\end{eqnarray}
Orbital coordinates in the frequency space take the form
\begin{eqnarray}\label{Eq:fourier_co_ord}
    x(\omega) = x^{*}(\omega) = \cfrac{a}{n}\,J_{n}^{\prime}(ne), \quad\qquad y(\omega) = -y^{*}(\omega) = \cfrac{\iota a}{n}\,\sqrt{\cfrac{1-e^2}{e^2}}\,J_{n}(ne).
\end{eqnarray}
Substituting the expression of Fourier space coordinates into Eq.~\eqref{Eq:ang_mom_scalar}, we obtain the following expression
\begin{eqnarray}\label{dJS}
    \cfrac{dJ^{i}_{s}}{dt} &=& -\cfrac{g_{s}^2}{6\pi} \left(\cfrac{N_{1}}{m_{1}}-\cfrac{N_{2}}{m_{2}}\right)^2 \mu^2 \omega_{0}^2 \left(1- \cfrac{n_{0S}^2}{n^2}\right)^{3/2} 2 x(\omega) (-\iota \omega) y(\omega) 
    \nonumber\\ &=&-\cfrac{g_{s}^2}{6\pi} \left(\cfrac{N_{1}}{m_{1}}-\cfrac{N_{2}}{m_{2}}\right)^2 \mu^2 a^2 \omega_{0}^3   \sum_{n\,\geq\,n_{0S}}^{\infty} \sqrt{\cfrac{1-e^2}{e^2}} \left(1- \cfrac{n_{0S}^2}{n^2}\right)^{3/2}2n\,J_{n}(ne)\,J_{n}^{\prime}(ne).
\end{eqnarray}
The rate of angular momentum loss resulting from the emission of a massless scalar
\begin{eqnarray}
    \cfrac{dJ_{s}^{i}}{dt}\Bigg|_{m_{\phi}=0}&=&-\cfrac{g_{s}^2}{6\pi}\,\left(\cfrac{N_{1}}{m_{1}}-\cfrac{N_{2}}{m_{2}}\right)^2\,\mu^2\,a^2\,\omega_{0}^3\,\sqrt{\cfrac{1-e^2}{e^2}}\,\sum_{n=1}^{\infty}2n\,J_{n}(ne)\,J_{n}^{\prime}(ne)\nonumber \\
    &=&- \cfrac{g_{s}^2}{12\pi} \left(\cfrac{N_{1}}{m_{1}}-\cfrac{N_{2}}{m_{2}}\right)^2 \mu^2 a^2 \omega_{0}^3\,\cfrac{1}{(1-e^2)}\,.
\end{eqnarray}
We can write the ratio of $|dJ_{s}^{i}/dt|$ and $dN_{s}/dt$ due to a massless scalar particle radiation for individual modes as
\begin{eqnarray}
    \mathcal{R}_{\rm Scalar} (n, e) = \hbar\, \cfrac{\sqrt{\cfrac{1-e^2}{e^2}}\,2 n J_{n}(ne) J_{n}^{\prime}(ne)}{n\left[J_{n}^{\prime 2}(ne) + \left(\cfrac{1-e^2}{e^2}\right)\,J_{n}^{2}(ne)\right]}\,.
\end{eqnarray}
Accordingly, we can estimate the ratio of the rate of angular momentum loss and the rate of number density loss as 
\begin{eqnarray}\label{Eq:ratio_j_n_scalar}
    \cfrac{|\dot{J}|}{\dot{N}}=\hbar\, \cfrac{\sqrt{\cfrac{1-e^2}{e^2}} \sum_{n\,=\,1}^{\infty} 2 n J_{n}(ne) J_{n}^{\prime}(ne)}{\sum_{n\,=\,1}^{\infty} n\left[J_{n}^{\prime 2}(ne) + \left(\cfrac{1-e^2}{e^2}\right)\,J_{n}^{2}(ne)\right]}\simeq \hbar\,.
\end{eqnarray}
\begin{figure}[htb!]
    \centering
    \includegraphics[width=0.495\linewidth]{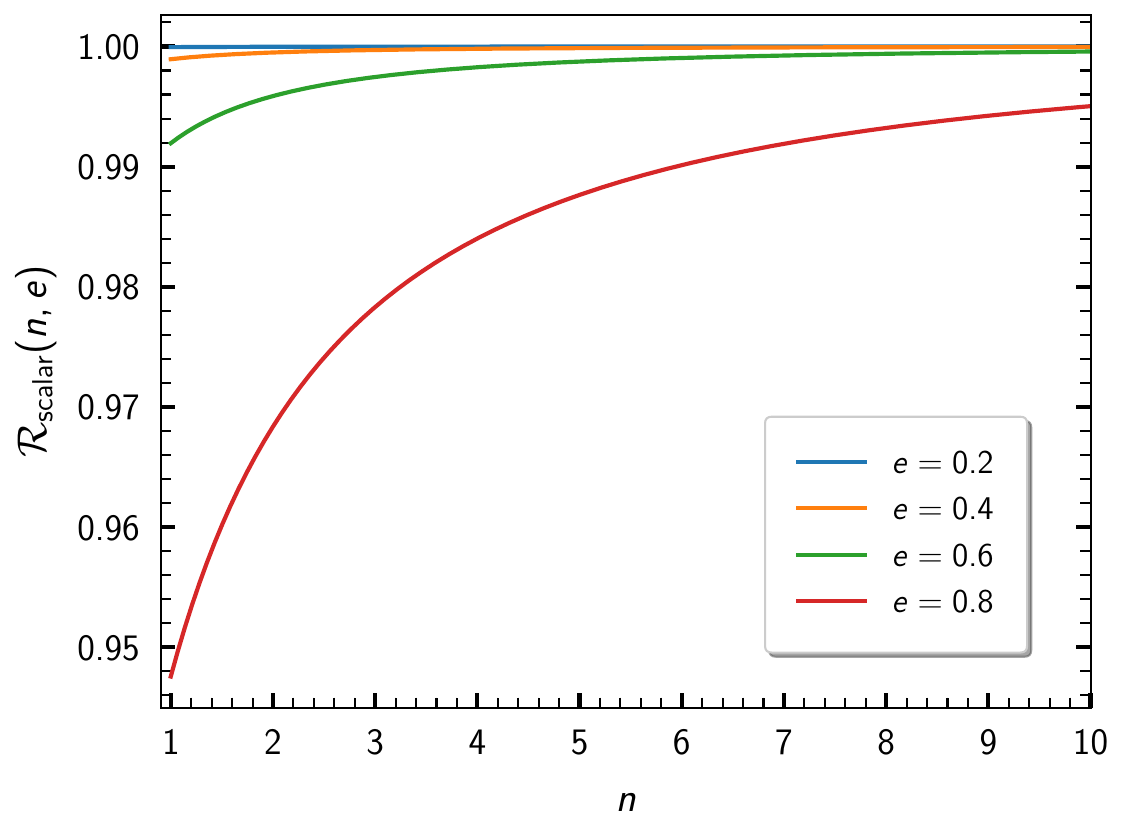}
    \includegraphics[width=0.495\linewidth]{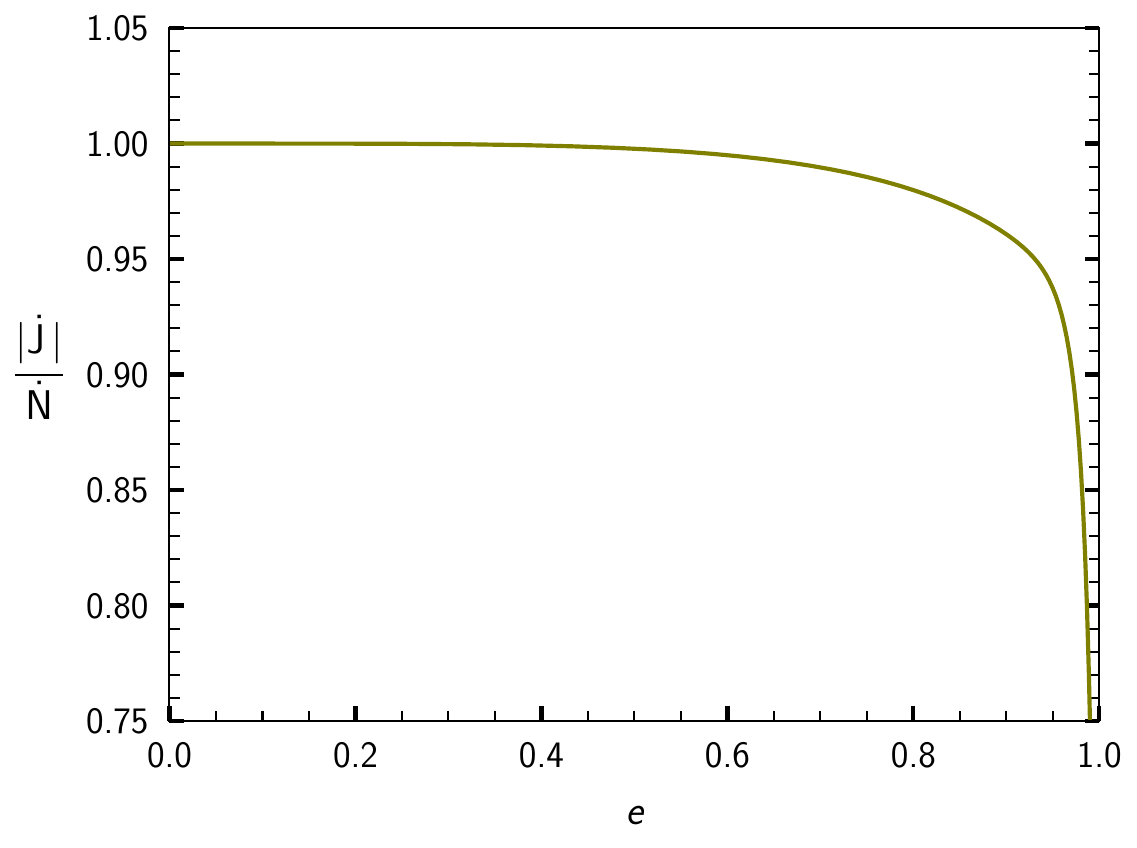}
    \caption{The ratio of the angular momentum loss rate to the scalar particle number emission rate, expressed in units of $\hbar$, is plotted as a function of the frequency mode for different orbital eccentricities in the left panel. The ratio starts from a small value at lower modes, increases with the mode number, and approaches $\hbar$ at higher modes. The right panel depicts the variation of $\frac{|\dot{J}|}{\dot{N}}$ (in units of $\hbar$) with orbital eccentricity. The ratio remains equal to $\hbar$ for $e < 1$, decreases with increasing eccentricity, and approaches $0$ in the limit $e \to 1$, as $|\dot{J}|$ vanishes in this limit.}
    \label{fig:ratio_nu_e_scalar}
\end{figure}

\noindent The ratio of the angular momentum loss rate to the scalar particle number emission rate, expressed in units of $\hbar$, is plotted as a function of the frequency mode for different orbital eccentricities in the left panel of Fig.~\ref{fig:ratio_nu_e_scalar}. The ratio starts from a small value at lower modes, increases with the mode number, and approaches $\hbar$ at higher modes. The right panel gives the corresponding ratio after summing over all modes. It remains equal to $\hbar$ for $e < 1$, and decreases with increasing eccentricity. The ratio approaches zero in the limit $e \to 1$, as $|\dot{J}|$ vanishes in this limit.


\subsection{Radiation of vector bosons from binary stars}

Neutron star binaries can also radiate vector bosons, provided that the mass of the vector boson lies below the orbital frequency of the binary. Ultralight vector bosons, which couple to the baryon and/or lepton number, can be radiated from compact binary star orbits and will contribute to the energy loss of the bound state, and can be constrained from the observations of period loss of binary pulsars~\cite{KumarPoddar:2019ceq}. An ultralight gauge boson $Z^\prime_\mu$ interacts with a fermionic current $J^\mu$ through the following Lagrangian
\begin{eqnarray}\label{eq:vector-lagrangian}
    \mathcal{L}_{V} =-\frac{1}{4}F_{\mu \nu}F^{\mu \nu} + \cfrac{M^2_{Z^\prime}}{2}\,Z^\prime_{\mu}Z^{\prime\mu} + gJ^{\mu}Z^\prime_{\mu}.
\end{eqnarray}
Spontaneous breaking of a gauge symmetry generates the mass term for the gauge boson. The energy loss rate associated with the vector boson radiation by a classical current is given as
\begin{eqnarray}\label{eq:rate-of-energy-loss-vector}
    \cfrac{dE_{V}}{dt} &=& g_{V}^{2} \int  \sum\limits_{\lambda\,=\,1}^{3}\,\left[ J^{\mu}(k^\prime)J^{\nu *}(k^\prime)\epsilon^{\lambda}_{\mu}(k)\epsilon^{\lambda*}_{\nu}(k) \right] \omega (2\pi) \delta(\omega - \omega^{\prime})\cfrac{d^{3}k}{(2\pi)^{3}\,2\omega}\,.
\end{eqnarray}
$J^{\mu}(k^\prime)$ is the Fourier space representation of the classical current $J^{\mu}(x)$. $\epsilon^{\lambda}_{\mu}(k)$ corresponds to the polarization vector of the massive vector boson, having four momentum $k$ and polarization index $\lambda$. The completeness relation for the polarization vector reads as
\begin{eqnarray}\label{eq:polarisation-sum}
\sum\limits_{\lambda\,=\,1}^{3}\epsilon^{\lambda}_{\mu}(k)\epsilon^{\lambda*}_{\nu}(k) = -g_{\mu\nu} + \cfrac{k_{\mu} k_{\nu}}{M^2_{Z^\prime}}\,.
\end{eqnarray}
The current density corresponding to the binary stars can be represented as
\begin{eqnarray}\label{eq:current-density}
    J^{\mu}(x) = \sum\limits_{a\, =\, 1,2} Q_{a}\delta^{3} (\mathbf{x}-\mathbf{x}_a(t))u^{\mu}_a.
\end{eqnarray}
Since the gauge bosons get generated from the spontaneous breaking of a gauge symmetry, the associated current density remains conserved, i.e., $\partial_\mu J^{\mu}(x)=0$. This constraint is imposed on the expression $|J^\mu J_\mu^*|$ while computing the energy and angular momentum loss rate. The quantity $Q_a$ corresponds to the total charge of $a$-th neutron star of the binary arising from its muon content, $\mathbf{x}_a(t)$ parametrizes the orbit of the binaries and $u^{\mu}_a = (1, \dot{x}_a, \dot{y}_a, 0)$ denotes the non-relativistic four-velocity in the $x$-$y$ plane of the binary. The spatial component of the current density admits the following representation in Fourier space:
\begin{eqnarray}\label{eq:current-density-fourier-full}
    J^i(\omega^\prime) = \sum\limits_{a\, =\, 1,2} \int \cfrac{1}{T} \int^T_0 dt\, e^{\iota\omega^{\prime} t}\,d^3\mathbf{x}\, e^{\iota \mathbf{k}^\prime \cdot \mathbf{x}}\, Q_{a} \delta^3 (\mathbf{x}-\mathbf{x}_a(t))\dot{x}_{a}(t)
    = \sum\limits_{a\, =\, 1,2} \cfrac{Q_{a}}{T} \int^T_0 dt \, e^{\iota\omega^{\prime} t} e^{\iota \mathbf{k}^\prime \cdot \mathbf{x}_{a}} \dot{x}_{a}(t).
\end{eqnarray}
With $\omega^\prime = n\omega_0$. For slowly moving sources $\mathbf{k}^{\prime}\cdot\mathbf{x}_{a} \ll \omega_0\, t$, accordingly we can perform a series expansion of $e^{\iota \mathbf{k}^{\prime} \cdot\mathbf{x}_{a}} = 1 + \iota \mathbf{k}^{\prime} \cdot\mathbf{x}_{a} + \dots$. Retaining only the leading term, the above expression reduces to
\begin{eqnarray}\label{eq:current-density-fourier-lo}
J^i(\omega^\prime) = \cfrac{Q_{1}}{T} \int^{T}_{0} dt \, e^{\iota \omega^{\prime} t} \dot{x}^{i}_{1} (t) + \cfrac{Q_{2}}{T} \int^{T}_{0} dt \, e^{\iota \omega^{\prime} t} \dot{x}^{i}_{2}(t). 
\end{eqnarray}
Recasting the coordinates in terms of the center of mass coordinates, using Eq.~\eqref{eq:com-coordinates}, and incorporating the rotation of the binary plane, as described in Eq.~\eqref{eq:coordinate-rotation}, we can identify the current density components as
\begin{eqnarray}\label{eq:current-density-com-rotated}
    J^{i}(\omega^\prime) &=& \left(\cfrac{Q_{1}}{m_{1}} - \cfrac{Q_{2}}{m_{2}}\right) \cfrac{\mu}{T} \int^{T}_{0} dt \, e^{\iota  \omega^{\prime} t} \mathcal{R}^{i}{}_{j}\dot{x}^{j}(t). 
\end{eqnarray}
The components are finally expressed in terms of the Bessel functions by substituting the Fourier transformed velocities from Eq.~\eqref{eq:fourier-coord-x-y},
\begin{eqnarray}\label{eq:current-density-x-y-z}
    J^{x}(\omega^{\prime}) = -\iota \tilde{Q} J^{\prime}_{n} (ne), \qquad
    J^{y}(\omega^{\prime}) = \sqrt{\cfrac{1-e^2}{e^2}}\,\tilde{Q}J_{n}(ne)\cos {i}, \qquad 
     J^{z}(\omega^{\prime}) =  \sqrt{\cfrac{1-e^2}{e^2}}\,\tilde{Q}J_n (ne) \sin i\,.
\end{eqnarray}
Here, $\tilde{Q} = \mu a\omega_{0}\left(\cfrac{Q_{1}}{m_{1}} - \cfrac{Q_{2}}{m_{2}}\right)$. This implies,
\begin{eqnarray}\label{eq:current-density-sq-spatial}
    J^{i}J^*_{i} = J^{x}J^{x*} + J^{y}J^{y*} + J^{z}J^{z*}
    = \tilde{Q}^2 \,\left[ J^{\prime\,2}_n (ne) + \cfrac{(1-e^2)}{e^2}\,J^2_n (ne) \right].
\end{eqnarray}
using Eq.~\eqref{eq:current-density}, we express the $0$-th component of the source current $J^{\mu}$ as  
\begin{eqnarray}\label{eq:current-density-sq-temporal-1_gauge}
    J^{0}(\omega^{\prime}) &=& \cfrac{1}{2\pi}\int dt\, d^{3}x^{\prime}\, e^{-\iota \omega^{\prime} t}  e^{\iota k^{\prime}\cdot x^{\prime}}\sum_{a\,=\,1,2}Q_{a}\delta^{3}(x^{\prime}-x_{a}(t)) \nonumber \\
    &=& \left(Q_{1}+Q_{2}\right)\delta^{(3)}(\omega^{\prime})+\iota \mu \left(\frac{Q_{1}}{m_{1}} - \frac{Q_{2}}{m_{2}}\right)|k^{\prime}|\left(\mathcal{F}_{1}(\theta^{\prime}, \phi^{\prime}) x(\omega^{\prime}) + \mathcal{F}_{2}(\theta^{\prime}, \phi^{\prime}) y(\omega^{\prime})\right).
\end{eqnarray}
We truncate the series expansion of $e^{\iota k^{\prime}\cdot x^{\prime}}$ at $\mathcal{O}(k^{\prime})$, which is justified as long as the source moves non-relativistically. The radiated power per unit solid angle arising from the emission of the vector boson simplifies to 
\begin{eqnarray}\label{eq:rate-of-energy-loss-vector-2}
    \cfrac{dE_{V}}{dt d\Omega} &=& \cfrac{g_{V}^{2}}{8\pi^2} \int d\omega \left[ -\left(J^{\mu}(\omega^{\prime})\right)^{2} + \cfrac{1}{M_{Z^{\prime}}^{2}}\left(\omega^{2} |J^0(\omega^\prime)|^{2} + 2 k_{0} k_{i} J^{0}(\omega^{\prime}) J^{i*}(\omega^{\prime}) +  k_{i} k_{j} J^{i}(\omega^{\prime}) J^{j*}(\omega^{\prime})\right) \right]  \nonumber\\
    &&\qquad \qquad  \times\, \omega^{2}\left(1 - \cfrac{M^2_{Z^\prime}}{\omega^2}\right)^{\frac{1}{2}} \delta(\omega - \omega^\prime).
\end{eqnarray}
We have,
\begin{eqnarray}\label{eq:current-density-sq-temporal-2_gauge}
        &&|J^0(\omega^{\prime})|^2 = \tilde{Q}^{2}\left(1- \cfrac{m_{Z^{\prime}}^{2}}{\omega^{\prime 2}}\right)
         \left[\mathcal{F}_{1} (\theta^\prime, \phi^\prime)^{2}  J^{\prime 2}_{n}(ne) + \mathcal{F}_{2} (\theta^\prime, \phi^\prime)^{2} \left(\cfrac{1-e^{2}}{e^{2}}\right) J^{2}_{n}(ne)  \right], \nonumber \\
       && |J^i(\omega^{\prime})|^2 = \tilde{Q}^{2}
     \left[J^{\prime 2}_{n}(ne) + \left(\cfrac{1-e^{2}}{e^{2}}\right) J^{2}_{n}(ne)  \right], \nonumber \\
     &&|J^0(\omega^{\prime})(\hat{n}^{\prime}\cdot \vec{J}(\omega^{\prime})| = \tilde{Q}^{2}\sqrt{\left( 1- \cfrac{m_{Z^{\prime}}^{2}}{\omega^{\prime 2}}\right)} \left[\mathcal{F}_{1} (\theta^\prime, \phi^\prime)^{2} J^{\prime 2}_{n}(ne) + \mathcal{F}_{2} (\theta^\prime, \phi^\prime)^{2} \left(\cfrac{1-e^{2}}{e^{2}}\right)J^{2}_{n}(ne)  \right], \nonumber \\
        &&|(\hat{n}^{\prime}\cdot \vec{J}(\omega^{\prime}))|^{2} = \tilde{Q}^{2}
        \left[\mathcal{F}_{1} (\theta^\prime, \phi^\prime)^{2}  J^{\prime 2}_{n}(ne) + \mathcal{F}_{2} (\theta^\prime, \phi^\prime)^{2} \left(\cfrac{1-e^{2}}{e^{2}}\right) J^{2}_{n}(ne) \right].
\end{eqnarray}
By substituting the expressions in Eq.~\eqref{eq:current-density-sq-temporal-2_gauge}, the quantity inside the square bracket in Eq.~\eqref{eq:rate-of-energy-loss-vector-2} can be written as
\begin{eqnarray}\label{eq:current-density-vector-3_gauge}
    &&\left[-\Big|J^{0}(\omega^{\prime})\Big|^2 + \Big|J^{i}(\omega^{\prime})\Big|^2 + \cfrac{\omega^{\prime 2}}{M_{Z^{\prime}}^{2}}\left(\Big|J^{0}(\omega^{\prime})\Big|^2 - 2 \sqrt{\left( 1- \cfrac{m_{Z^{\prime}}^{2}}{\omega^{\prime 2}}\right)}\,\Big|J^0(\omega^{\prime})(\hat{n}^{\prime}\cdot \vec{J}(\omega^{\prime})\Big|
   + \left( 1- \cfrac{m_{Z^{\prime}}^{2}}{\omega^{\prime 2}}\right)\Big|(\hat{n}^{\prime}\cdot \vec{J}(\omega^{\prime})\Big|^{2}\right) \right] \nonumber \\
    && = \tilde{Q}^{2}\Bigg[\Bigg(1 + \Bigg(-1+ \cfrac{M_{Z^{\prime}}^{2}}{\omega^{\prime 2}}  \Bigg)\mathcal{F}_{1} (\theta^\prime, \phi^\prime)^2 \Bigg) J^{\prime 2}_{n}(ne) + \Bigg(1 + \Bigg(-1+ \cfrac{M_{Z^{\prime}}^{2}}{\omega^{\prime 2}}  \Bigg)\mathcal{F}_{2} (\theta^\prime, \phi^\prime)^2 \Bigg) \left(\cfrac{1-e^{2}}{e^{2}}\right) J^{2}_{n}(ne) \Bigg] \nonumber \\
   && = \tilde{Q}^{2} \mathcal{F}(n, e, \theta^{\prime}, \phi^{\prime}).
\end{eqnarray}
The angular functions $\mathcal{F}_{1,2}(\theta^\prime, \phi^\prime)$ are defined in Eq.~\eqref{eq:x-y-coeff-angular}. Using the above expression in the flux of energy radiated owing to the vector emission (Eq.~\eqref{eq:rate-of-energy-loss-vector-2}), we find
\begin{eqnarray}\label{eq:rate-of energy-loss-vector-angular}
    \cfrac{dE_{V}}{dt d\Omega} &=& \cfrac{g_{V}^{2}}{8\pi^{2}}\,\mu^2  a^2 \omega_{0}^{4}  \left(\cfrac{Q_{1}}{m_{1}} - \cfrac{Q_{2}}{m_{2}}\right)^2 \sum_{n\,\geq\,n_{0V}}^{\infty}  n^2  \left(1 - \cfrac{n_{0V}^{2}}{n^{2}}\right)^{\frac{1}{2}}  \mathcal{F}(n, e, \theta^{\prime}, \phi^{\prime}),
\end{eqnarray}
where, $n_{0V}=M_{Z^{\prime}}/\omega_{0}$ is the harmonic of the fundamental frequency $\omega_0$, which, in the presence of
vector mediated fifth force, is given by $\omega_0=\sqrt{G(1+\alpha)M/a^3}$, with  $\alpha = g_{V}^2 Q_{1} Q_{2}/(4\pi Gm_{1} m_{2})$ (when the orbital separation is such that we have $m_{V} r \ll 1$).
After performing the angular integration, we obtain
\begin{eqnarray}\label{dEV}
    \cfrac{dE_{V}}{dt} = \cfrac{g_{V}^{2}}{6\pi}\, \mu^2 a^2 \omega_{0}^4 \left(\cfrac{Q_{1}}{m_{1}}-\cfrac{Q_{2}}{m_{2}}\right)^2\sum_{n\,\geq\,n_{0V}}^{\infty} n^{2} \left(1-\cfrac{n_{0V}^{2}}{n^2}\right)^{1/2} \left(2+\cfrac{n_{0V}^{2}}{n^2}\right) \left[J_{n}^{\prime 2}(ne) + \left(\cfrac{1-e^2}{e^2}\right) J_{n}^{2}(ne)\right].
\end{eqnarray}
Similarly, the number of vector quanta emitted is obtained by dividing each term in the energy mode sum in Eq.~\eqref{dEV} by $n \hbar \omega_0 $. The resulting expression is
\begin{eqnarray} \label{dNV}
    \cfrac{dN_{V}}{dt}
    &=& \cfrac{g_{V}^{2}}{6\pi\hbar}\, \mu^{2} a^{2} \omega_{0}^{3} \left(\cfrac{Q_{1}}{m_{1}}-\cfrac{Q_{2}}{m_{2}}\right)^{2} \nonumber \\ && \times \sum_{n\,\geq\,n_{0V}}^{\infty} n \left(1-\cfrac{n_{0V}^2}{n^2}\right)^{1/2}\left(2+\cfrac{n_{0V}^2}{n^2}\right) \left[J_{n}^{\prime 2}(ne) + \left(\cfrac{1-e^2}{e^2}\right)\,J_{n}^2(ne)\right].
\end{eqnarray}
In the massless limit ($m_{Z^{\prime}} \to 0$), the above equation reduces to
\begin{eqnarray}
    \cfrac{dN_{V}}{dt}\Bigg|_{m_{Z^{\prime}}=0}=\cfrac{g_{V}^{2}}{6\pi\hbar}\,\mu^{2} a^{2} \omega_{0}^{3}\left(\cfrac{Q_{1}}{m_{1}}-\cfrac{Q_{2}}{m_{2}}\right)^2 \sum_{n\,=\,1}^{\infty} 2n\left[J_{n}^{\prime 2}(ne) + \left(\cfrac{1-e^2}{e^2}\right) J_{n}^2(ne)\right].
\end{eqnarray}
The angular momentum loss resulting from the vector radiation is given by 
\begin{eqnarray}\label{Eq:ang_mom_vector}
    \cfrac{dJ_{V}^{i}}{dt} &=& \cfrac{g_{V}^2}{6\pi}\left(\cfrac{Q_{1}}{m_{1}} - \cfrac{Q_{2}}{m_{2}}\right)^{2} \mu^{2} \int \epsilon^{ijk} x_{j} p_{k} \left(p_{l} p_{l}^{*} - p_{0} p_{0}^{*}\right) \left(2+\cfrac{M_{Z^{\prime}}^2}{\omega^{\prime 2}}\right)\left(1-\cfrac{M_{Z^{\prime}}^2}{\omega^{\prime 2}}\right)^{1/2} \omega\,\delta(\omega -\omega^{\prime}) d\omega\nonumber \\
    &=& \cfrac{g_{V}^2}{6\pi}\left(\cfrac{Q_{1}}{m_{1}} - \cfrac{Q_{2}}{m_{2}}\right)^2\iota\mu^2 \int \epsilon^{ijk} p_{k} p_{j}^{*} \left(2+\cfrac{M_{Z^{\prime}}^2}{\omega^{\prime 2}}\right) \left(1-\cfrac{M_{Z^{\prime}}^2}{\omega^{\prime 2}}\right)^{1/2} \omega\,\delta(\omega -\omega^{\prime}) d\omega\nonumber \\
    &=& -\cfrac{g_{V}^2}{6\pi}\left(\cfrac{Q_{1}}{m_{1}} - \cfrac{Q_{2}}{m_{2}}\right)^2\iota\mu^{2} \int \left(\dot{x}(\omega^{\prime}) \dot{y}^{*}(\omega^{\prime}) - \dot{x}^{*}(\omega^{\prime}) \dot{y}(\omega^{\prime}) \right)\left(2+\cfrac{M_{Z^{\prime}}^2}{\omega^{\prime 2}}\right)\left(1-\cfrac{M_{Z^{\prime}}^2}{\omega^{\prime 2}}\right)^{1/2} \omega\,\delta(\omega -\omega^{\prime}) d\omega.
\end{eqnarray}
The Fourier transform of the time derivatives of the orbit coordinates is read as follows
\begin{eqnarray}
    \dot{x}(\omega) = - \dot{x}^{*}(\omega) = -\iota a\omega_{0} J_{n}^{\prime}(ne), \qquad\quad \dot{y}(\omega) = \dot{y}^{*}(\omega) = \sqrt{\cfrac{1-e^2}{e^2}}\,a\omega_{0} J_{n}(ne)
\end{eqnarray}
Inserting the above expressions into Eq.~\eqref{Eq:ang_mom_vector} yields the following angular momentum loss rate
\begin{eqnarray} \label{dJV}
    \cfrac{dJ^{i}_{V}}{dt} &=& \cfrac{g_{V}^2}{6\pi}\left(\cfrac{Q_{1}}{m_{1}} - \cfrac{Q_{2}}{m_{2}}\right)^2 \mu^{2} \left(2+\cfrac{n_{0V}^2}{n^2}\right) \left(1-\cfrac{n_{0V}^2}{n^2}\right)^{1/2} \dot{x}(\omega) (-\iota \omega) \dot{y}(\omega) \nonumber \\
    &=& -\cfrac{g_{V}^2}{6\pi}\left(\cfrac{Q_{1}}{m_{1}} - \cfrac{Q_{2}}{m_{2}}\right)^{2} \mu^{2} a^{2} \omega_{0}^{3} \sum_{n\,\geq\,n_{0V}}^{\infty} \left(2+\cfrac{n_{0V}^2}{n^2}\right)\left(1-\cfrac{n_{0V}^2}{n^2}\right)^{1/2}\sqrt{\cfrac{1-e^2}{e^2}}\,n J_{n}(ne) J_{n}^{\prime}(ne).
\end{eqnarray}
In the massless limit, this reduces to 
\begin{eqnarray}
    \cfrac{dJ_{V}^{i}}{dt}\Bigg|_{m_{Z^{\prime}}=0} &=& -\cfrac{g_{V}^2}{6\pi} \left(\cfrac{Q_{1}}{m_{1}} - \cfrac{Q_{2}}{m_{2}}\right)^{2} \mu^{2} a^{2} \omega_{0}^{3} \sum_{n\,=\,1}^{\infty} \sqrt{\cfrac{1-e^2}{e^2}}\,2n J_{n}(ne) J_{n}^{\prime}(ne)\nonumber \\
    &=& -\cfrac{g_{V}^2}{12\pi} \left(\cfrac{Q_{1}}{m_{1}} - \cfrac{Q_{2}}{m_{2}}\right)^{2} \mu^{2} a^{2} \omega_{0}^{3}\,\cfrac{1}{(1-e^2)}.
\end{eqnarray}
As in the case of scalar emission, the ratio of $|dJ_{V}^{i}/dt|$ and $dN_{V}/dt$ for a massless vector particle emission for individual modes is given by
\begin{eqnarray}
    \mathcal{R}_{\rm Vector} (n, e) = \hbar\, \cfrac{\sqrt{\cfrac{1-e^2}{e^2}}\,2n J_{n}(ne) J_{n}^{\prime}(ne)}{n\left[J_{n}^{\prime 2}(ne) + \left(\cfrac{1-e^2}{e^2}\right) J_{n}^{2}(ne)\right]}\,.
\end{eqnarray}
Accordingly, we can estimate the ratio of the rate of angular momentum loss and the rate of number density loss as 
\begin{eqnarray}\label{Eq:ratio_j_n_vector}
    \cfrac{\dot{J}}{\dot{N}}=\hbar\, \cfrac{\sqrt{\cfrac{1-e^2}{e^2}} \sum_{n\,=\,1}^{\infty} 2 n J_{n}(ne) J_{n}^{\prime}(ne)}{\sum_{n\,=\,1}^{\infty}\,n\left[J_{n}^{\prime 2}(ne) + \left(\cfrac{1-e^2}{e^2}\right) J_{n}^{2}(ne)\right]}\simeq \hbar\,.
\end{eqnarray}

\section{Determination of spin of the radiated field from binary observations}\label{Sec:spin_determination}

In this section, we elaborate on how the loss of energy and angular momentum changes the parameters of the Kepler orbit~\cite{Seymour:2020yle, Chen:2025mwl}. We consider a compact binary system bound by gravity and a possible fifth force, mediated by scalar or vector particle exchange. The binding energy and angular momentum of a binary in a Keplerian orbit mediated by gravity and a fifth force are given by
\begin{equation}
    E = -\cfrac{\mu(1+\alpha) G   M}{2a}\,,\quad J = [\mu^2(1+\alpha)a GM (1-e^2)]^{1/2}\,.
\end{equation}
When the fifth force is mediated by a vector exchange, the corresponding $\alpha = g_{V}^2 Q_{1} Q_{2}/(4\pi Gm_{1} m_{2})$, provided that the orbital separation satisfies $m_{V} r \ll 1$. In the case of scalar-mediated fifth-force interaction, $\alpha=- g_s^2 N_1 N_2/(4 \pi G m_1 m_2)$, when the orbital separation satisfies $m_s r \ll 1$.  As the binary system loses energy, the binding energy becomes more negative, the semi-major axis $a$ decreases, and the angular frequency $\omega_0 = \cfrac{2\pi}{T} = \left( \cfrac{(1+\alpha)GM}{a^3} \right)^{1/2}$ decreases at rates which are related to the energy loss as
\begin{equation}\label{dEdt-orbit}
 \cfrac{dE}{dt} = \cfrac{\mu (1+\alpha)GM}{2a^2}\, \dot{a} = -\cfrac{\mu}{3} \left[(1+\alpha)aGM \right]^{1/2}\, \dot{\omega}_{0}\,.
\end{equation}
The loss of angular momentum results in both an increase in the angular frequency $\omega_0$ of the orbit (which is related to the change in $a$) and a decrease in eccentricity $e$ of the Kepler orbit, given by
\begin{equation}\label{dJdt-orbit}
    \frac{dJ}{dt} = -(1-e^2)^{1/2}\,\cfrac{\mu}{3}\,a^2\dot{\omega}_{0} - \dot{e}\cfrac{e}{(1-e^2)^{1/2}}\,\mu\left[(1+\alpha)GM\right]^{1/2}a^{1/2}.
\end{equation}
The energy and angular momentum carried away in gravitational, and possibly scalar and vector radiation, result in a decrease in the orbital energy and orbital angular momentum
\begin{equation}\label{extraEJ}
   - \frac{dE}{dt}= \frac{dE}{dt}\Big\vert_{GW} + \frac{dE}{dt}\Big\vert_{S,V} \,, \quad - \frac{dJ}{dt}= \frac{dJ}{dt}\Big\vert_{GW} + \frac{dJ}{dt}\Big\vert_{S,V}\,.
\end{equation}
The energy flux carried by gravitational waves is given by the Peters-Mathews formula in Eq.~\eqref{PWE}. We derived this result in the frequency space using the field-theoretic method. We use the same approach to derive the energy flux carried by ultralight scalars and dark-photon vector radiation, given in Eqs.~\eqref{dES} and \eqref{dEV}, respectively. The angular momentum carried by gravitational waves was first calculated by Peters using classical GR~\cite{Peters:1964zz}. In this paper, we derived the same result in the frequency-space formalism, given in Eq.~\eqref{dGJ}. The corresponding angular momentum fluxes carried by scalars and vectors are given in Eqs.~\eqref{Eq:ang_mom_scalar} and \eqref{dJV}, respectively. The evolution of orbital parameters $\dot{a}, \dot {\omega}_{0}, \dot {e}$ provides a direct measure of the energy and angular momentum losses of the binary system through Eqs.~\eqref{dEdt-orbit} and \eqref{dJdt-orbit}. Any discrepancy between the observations and the predictions of GR can provide observational evidence for the additional energy and angular momentum losses encoded in Eq.~\eqref{extraEJ}.


\section{CONCLUSIONS}\label{Sec:conclusions}

Gravitational waves carry orbital and spin angular momentum of the graviton. Direct detection of GWs may probe the angular momentum component of the GWs~\cite{Baral:2019dke}. Another way to probe the angular momentum carried away by GWs is through the observations of binary pulsars of the Hulse-Taylor~\cite{Hulse:1974eb} type. In this paper, we computed the angular momentum loss of binary orbits due to graviton, scalar, and dark-photon radiation. We perform a field-theoretic frequency-spectrum calculation of the energy and angular-momentum radiation for elliptical and hyperbolic orbits. 

For elliptical orbits, the gravitational wave energy and angular momentum are calculated for each harmonic of the fundamental frequency $\omega^{\prime}= n \omega_0$ with $n$ being a positive integer. While for hyperbolic orbits the frequency spectrum $\omega^\prime= \nu \omega_0$  with $\nu$ real and $\nu \in [0, \infty)$. We compute the energy and angular-momentum radiation rates of gravitons by starting with linearized gravity and using Fermi's golden rule. The rates of graviton emission from a classical source are given by the amplitude squared $|{\cal M}|^2=|\kappa T^{\mu \nu} h_{\mu \nu}|^2$, summed over the final state phase space. For binaries in an elliptical orbit, we reproduce the classical GR results~\cite{Peters:1963ux, Peters:1964zz}. Keeping track of all factors of $\hbar$, we find that the energy (Eq.~\eqref{PWE}) and angular momentum  (Eq.~\eqref{Eq:dJ_dt_grav_ellip}) radiation rates are independent of $\hbar$. In our method, we can separately compute the radiated orbital angular momentum (Eq.~\eqref{dLdt-4}) and spin angular momentum (Eq.~\eqref{dSdt-5}). We find that the orbital angular momentum per graviton $\dot{L}/\dot{N} \simeq (2/3) \hbar$, while the spin angular momentum per graviton is $\dot{S}/\dot{N} \simeq (4/3) \hbar$. Thus, the total angular momentum per graviton is $\dot{J}/\dot{N} \simeq 2 \hbar$. This ratio is also eccentricity dependent, as we show in Fig.~\ref{fig:Ang_Mom_ratio}. The splitting of the graviton's total angular momentum into spin and orbital parts is not gauge-independent. Only the total radiated angular momentum can be inferred from the observations of binary orbits. 
A recent paper by Page~\cite{Page:2025ncu} uses the classical calculations of energy radiation~\cite{Peters:1963ux} and the rate of angular momentum loss~\cite{Peters:1963ux} per orbit by elliptical binaries as gravitational waves, and showed that the ratio is $\dot{J}/\dot{E}= 2 /{\omega}_{0}$. Then, defining the graviton-number radiation rate per orbit as $\dot{N} \equiv \dot{E} /(n\hbar \omega_0)$, it was pointed out that the ratio $\dot{J}/\dot{N} \simeq 2 \hbar$. 
We compute the energy and angular momentum radiated in elliptical orbits for each mode $n$ of the angular frequency $\omega^\prime_n= n \omega_0$. In Fig.~\ref{fig:loss_rate_Energy_ellp}, we plot the energy spectrum radiated in each frequency mode. The angular momentum and graviton number in each mode are shown in Fig.~\ref{fig:dotJ_dotN_ellip}. Finally, in the bottom panel of Fig.~\ref{fig:dotJ_dotN_ellip}, we show the angular momentum loss per radiated graviton $\mathcal{R}_{\rm ell} (n, e)$ at each frequency mode. We see that this ratio deviates significantly from $2\hbar$ at lower $n$. This deviation from $2\hbar$ at low $n$ becomes more pronounced with increasing eccentricity.

We further calculate the gravitational-wave energy and angular momentum radiated during hyperbolic encounters. The spectrum is continuous $\omega^\prime= \nu \omega_0$ with $\nu \in [0,\infty)$. The $\omega^\prime=0$ limit is particularly interesting as this represents the memory signal of the gravitational waveform~\cite{Hait:2022ukn}. We plot the energy spectrum $\frac{d\tilde E(\omega^\prime, e)}{d\omega^\prime}$ in Fig.~\ref{fig:Edot_nu_hyp} (left panel) and the total energy radiated across
all frequencies as a function of the eccentricity of the hyperbolic orbit (right panel). We see that the energy spectrum is nonzero in the $\nu\rightarrow 0$ limit, corresponding to the energy carried by the memory signal. In Fig.~\ref{fig:Ndot_nu}, we present the graviton number spectrum. The angular momentum spectrum is depicted in Fig.~\ref{fig:Jdot_nu}. The angular momentum per graviton number $\mathcal{R}_{\rm hyp}(\nu, e)$ as a function of frequency is shown in Fig.~\ref{fig:ratio_nu_e_hyp}. From these plots, we see that in the $\nu \rightarrow 0$ limit, the gravitational angular momentum is nonzero, representing the angular momentum carried by the gravitational memory signal from hyperbolic orbits. The graviton number diverges in the $\nu\rightarrow 0$ limit and therefore the angular momentum per graviton ratio $\mathcal{R}_{\rm hyp}(\nu, e) \rightarrow 0$ in this limit. Finally, we plot the ratio of the total angular momentum to the gravitons radiated per orbit summed over all frequencies, $\Delta J/\Delta N$, in Fig.~\ref{DJDNHyp}. We see that this ratio is a function of $e$, and the ratio saturates to $2\hbar$ for large eccentricity. This is the extension of Page~\cite{Page:2025ncu} observation to the case of hyperbolic orbits.

For Hulse-Taylor type compact binaries, which are separated by distance $a\simeq 1\, {\rm A.U}$, having masses $m_i\simeq 1.4\, M_\odot$, the fundamental frequency turns out to be $\omega_0=\sqrt{GM/a^3} \simeq 10^{-20} {\rm eV}$. This is in the mass range of scalar and vector particles, which can constitute the ultralight dark matter ~\cite{Hu:2000ke, Hui:2016ltb, Schive:2025bcm}. The energy loss due to the additional scalar or vector particles has been studied using classical field equations~\cite{Krause:1994ar, Chen:2025mwl} or in a field-theoretic framework~\cite{Mohanty:1994yi, KumarPoddar:2019jxe, Poddar:2023pfj, KumarPoddar:2019ceq}. The goal is to use observations of binary orbits to constrain the properties of these extra fifth-force particles~\cite{Seymour:2020yle, Cheng:2023qys}. In this paper, we extended this framework to calculate the angular momentum radiated via scalar and vector fields. This allows us to explore whether observations can also be used to determine the spin of the additional radiated particles.

We compute the energy loss due to the ultralight scalar radiation from the elliptical binary and obtain the closed-form expression in Eq.~\eqref{dES}. The angular momentum carried away by scalar radiation is given in Eq.~\eqref{dJS}. For vector radiation, we derive the corresponding expressions for the energy and angular momentum losses in Eqs.~\eqref{dEV} and \eqref{dJV}, respectively. We find that apart from the couplings and the extra degrees of freedom associated with the vector fields, the expressions for the scalar and vector are identical in form. One can differentiate the scalar radiation from vector radiation by measuring  $\dot e, \dot \omega_{0}$ from binary orbital observations and by using Eqs.~\eqref{dEdt-orbit} $-$ \eqref{extraEJ}, which relate the energy and angular momentum losses to the orbital parameters and determine the sign of $\alpha$. A positive value of $\alpha$ from observations will provide evidence for a vector-mediated fifth force, whereas a negative value would indicate a scalar-mediated fifth force.

We compute the angular momentum carried by scalars per quantum as a function of the angular frequency, as shown in the left panel of Fig.~\ref{fig:ratio_nu_e_scalar}. This ratio is close to $\hbar$ and deviates from $\hbar$ at low frequencies. The angular momentum per emitted quantum, summed over all frequencies, is close to $\hbar$. It deviates from this value only at large eccentricities and vanishes in the limit $e \to 1$, as shown in the right panel of Fig.~\ref{fig:ratio_nu_e_scalar}. 

Since the expressions for energy and angular momentum radiated are identical for scalar and vector radiation, apart from the couplings, the angular momentum carried per quantum is also close to $\hbar$ for vector radiation. Fig.~ \ref{fig:ratio_nu_e_scalar} also represents the spectrum and eccentricity dependence of the angular momentum carried per emitted quantum of vector radiation.

\begin{table}[h!]
\centering
\begin{tabular}{||c || c || c || c||} 
 \hline
 Radiated Particles & Source Current & Orbit & $\dot{J}/\dot{N}$ \\ [0.5ex] 
 \hline\hline
Graviton & Quadrupolar & Elliptical & $2\hbar$ \\ \hline
Graviton & Quadrupolar & Hyperbolic & $2\hbar$ \\ \hline
Scalar   & Dipolar     & Elliptical & $\hbar$ \\ \hline
Vector   & Dipolar     & Elliptical & $\hbar$ \\ [1ex] 
 \hline\hline
\end{tabular}
\caption{The angular momentum per quantum of graviton, scalar, and vector radiation, emitted by binaries in elliptical orbits and hyperbolic encounters.}
\label{table:1}
\end{table}

In Table~\ref{table:1} we summarize the angular momentum per quantum of particles of different spins. This ratio is $2\hbar$ for gravitational waves radiated by binaries in elliptical orbits (as first pointed out in ref.~\cite{Page:2025ncu}) as well as for hyperbolic orbits. We also calculate the angular momentum per quantum for scalar and vector radiation from elliptic binaries and find that it is $\hbar$ for both cases. A common feature of scalar and vector radiation is that the source current originates from a change in dipole moment. In contrast, for gravitational radiation, the current is sourced by a change in the quadrupole moment. The angular momentum per quantum of bosons of different spins may be related to the topological property of the Poincaré group representation of massless bosons~\cite{Palmerduca:2023ctx, Palmerduca:2024msq}. Experimental verification of the relation $\dot {J}/\dot{N} \simeq 2 \hbar$ ratio for gravitational waves can then establish the spin-2 nature of gravitons. 

\section{Data Availability}
No observational or experimental data were created or analyzed in this study.


\begin{acknowledgments}
The authors acknowledge the ICTS Workshop ``Hearing beyond the standard model with cosmic sources of Gravitational Waves" (code: 41/2024-12-P274-BGW), where this problem was initiated. The authors thank Debtosh Chowdhury and Diptarka Das for helpful discussions and their comments on the manuscript. AH acknowledges the financial support from the Fellowship for Academic and Research Excellence (FARE) (FARE/2025-26/IITK/51), Indian Institute of Technology Kanpur, India.
\end{acknowledgments}


\section{Appendix}
\subsection{Fourier transform of elliptical  orbit coordinates}
\label{app:fourier-coordinates}
\noindent The elliptical coordinates in terms of the anomaly parameters
\begin{eqnarray}\label{eq:elliptic-coordinate_App}
\quad x(\xi) = a\,(\cos\xi - e),\quad y(\xi) = a \sqrt{1 - e^2}\,\sin\xi, \quad z(\xi) = 0,\quad \omega_0\, t = (\xi - e\sin\xi).
\end{eqnarray}
The Fourier transform of the coordinates
\begin{eqnarray}\label{eq:fourier-coord-x-y}
    x(\omega^\prime) &=& \cfrac{1}{T}\int_{0}^{T} dt\,e^{-\iota\omega^{\prime}t}\,x(t) =  \cfrac{1}{\iota n\omega_{0} T} \,\int^T_0 dt\,e^{-\iota n\omega_{0} t}\,\dot{x} (t) = -\cfrac{\iota}{2\pi n} \,\int_{0}^{2\pi} d\xi\,e^{-\iota n(\xi -e\sin{\xi})}\,\cfrac{dx}{d\xi} \nonumber \\ &=& \cfrac{\iota a}{2\pi n}\,\int^{2\pi}_0\,d\xi\,\sin\xi\,e^{\iota n \, \left(\xi - e\,\sin\xi \right)} 
    = \cfrac{a}{2n}\,\left[ J_{n-1}(ne) - J_{n+1}(ne) \right]  = \cfrac{a}{n}\,J^\prime_n (ne). \nonumber\\
    y (\omega^\prime) &=& \cfrac{1}{T}\int_{0}^{T} dt\,e^{-\iota\omega^{\prime}t}\,y(t) =  \cfrac{1}{\iota n\omega_{0} T} \,\int^T_0 dt\,e^{-\iota n\omega_{0} t}\,\dot{y} (t) = -\cfrac{\iota}{2\pi n} \,\int_{0}^{2\pi} d\xi\,e^{-\iota n(\xi -e\sin{\xi})}\,\cfrac{dy}{d\xi} \nonumber \\  &=&\cfrac{\iota a\,\sqrt{1-e^2}}{2\pi n }\,\int^{2\pi}_0 d\xi\,\cos\xi\,e^{\iota n \left(\xi - e\,\sin\xi \right)} = \sqrt{1-e^2}\,\cfrac{\iota a}{2n}\,\left[J_{n+1}(ne) + J_{n-1}(ne)\right]
    = \cfrac{\iota a}{n}\,\sqrt{\cfrac{1-e^2}{e^2}}\,J_n (ne).
\end{eqnarray}

\subsection{Bessel function sums} \label{Bessel-formulas}
Sum over Bessel functions \cite{Peters:1963ux}:
\begin{eqnarray}\label{eq:Peter-Matthews-sum}
&& \sum_{n\,=\,1}^{\infty}  n^{2}J_{n}^{2}(ne) = \cfrac{e^{2}}{4(1-e^{2})^{7/2}}\left(1 + \cfrac{e^{2}}{4} \right), \label{B1} \\
&& \sum_{n\,=\,1}^{\infty}  n^{2}[J_{n}^{\prime}(ne)]^{2} = \cfrac{1}{4(1-e^{2})^{5/2}}\left(1 + \cfrac{3e^{2}}{4} \right), \label{B2} \\
&& \sum_{n\,=\,1}^{\infty} n J_{n}(ne) J^{\prime}_{n}(ne) = \cfrac{e}{4\left(1-e^2\right)^{3/2}}\,,\label{B3} \\
&& \sum_{n\,=\,1}^{\infty}  n^{3} J_{n}(ne) J_{n}^{\prime}(ne) = \cfrac{e}{4(1-e^{2})^{9/2}}\left(1 + 3e^{2} + \cfrac{3}{8}\,e^{4} \right),  \label{B4}\\
&& \sum_{n\,=\,1}^{\infty}  n^{4} J_{n}^{2}(ne) = \cfrac{e^{2}}{4(1-e^{2})^{13/2}}\left(1 + \cfrac{37}{4}\,e^{2} + \cfrac{59}{8}\,e^{4} + \cfrac{27}{64}\,e^{6} \right), \label{B5} \\
&& \sum_{n\,=\,1}^{\infty}  n^{4} [J_{n}^{\prime}(ne)]^{2} = \cfrac{1}{4(1-e^{2})^{11/2}}\left(1 + \cfrac{39}{4}\,e^{2} + \cfrac{79}{8}\,e^{4} + \cfrac{45}{64}\,e^{6} \right). \label{B6}
\end{eqnarray}


\bibliographystyle{JHEP}
\bibliography{reference}

\end{document}